\documentclass[prb,twocolumn,amsmath,amssymb]{revtex4-1}

\usepackage{graphicx}
\usepackage{dcolumn}
\usepackage{bm}
\usepackage{verbatim}
\usepackage{setspace}
\usepackage{fancyhdr,lastpage}

\begin{document}

\newcommand{\fourqquads}{{\qquad \qquad \qquad \qquad}}
\newcommand{\eightqquads}{{\qquad \qquad \qquad \qquad \qquad \qquad \qquad \qquad}}
\newcommand{\fournegspc}{{\! \! \! \!}}
\newcommand{\eightnegspc}{{\! \! \! \! \! \! \! \!}}
\newcommand{\AGAff}{${\rm Al_{0.55}Ga_{0.45}As}$}
\newcommand{\AGAy}{${\rm Al_{\it y}Ga_{1-{\it y}}As}$}
\newcommand{\AGAx}{${\rm Al_{\it x}Ga_{1-{\it x}}As}$}
\newcommand{\AGAf}{${\rm Al_{0.15}Ga_{0.85}As}$}
\newcommand{\AGAt}{${\rm Al_{0.10}Ga_{0.90}As}$}
\newcommand{\IGAx}{${\rm In_{0.53}Ga_{0.47}As/In_{0.52}Al_{0.48}As}$}
\newcommand{\IGx}{${\rm In_{0.53}Ga_{0.47}As}$}
\newcommand{\AGA}{GaAs/AlGaAs}
\newcommand{\IGA}{InGaAs/InAlAs}
\newcommand{\WcmK}{W/cm$\cdot$K}
\newcommand{\E}[1]{$E_{#1}$}
\newcommand{\N}[1]{$N_{#1}$}
\newcommand{\DN}{$\Delta N_{32}$}
\newcommand{\taut}[2]{$\tau_{#1}^{#2}$}
\newcommand{\Schrodinger}{Schr\"{o}dinger}
\newcommand{\FPerot}{Fabry-P\'{e}rot}
\newcommand{\epmiwt}{{e^{\pm i \omega t}}}
\newcommand{\empiwt}{{e^{\mp i \omega t}}}
\newcommand{\eiwt}{{e^{i \omega t}}}
\newcommand{\epiwt}{{e^{+i \omega t}}}
\newcommand{\emiwt}{{e^{-i \omega t}}}
\newcommand{\tauul}{{\tau_{u\rightarrow l}}}
\newcommand{\hw}{{\hbar \omega}}
\newcommand{\hwz}{{\hbar \omega_0}}
\newcommand{\kB}{{k_\mathrm{B}}}
\newcommand{\cc}{{\mathrm{c.c.}}}
\newcommand{\Ef}{{\mathcal{E}}}
\newcommand{\Pol}{{\mathcal{P}}}
\newcommand{\Rdiff}{{\mathcal{R}}}
\newcommand{\Rdiffm}{{\mathcal{R}_\mathrm{th}^-}}
\newcommand{\Rdiffp}{{\mathcal{R}_\mathrm{th}^+}}
\newcommand{\dw}{{d \omega}}
\newcommand{\hwab}{{\hbar \omega_{a b}}}
\newcommand{\Np}{{N_\mathrm{p}}}
\newcommand{\Pout}{{P_\mathrm{out}}}
\newcommand{\GammabyNp}{{\frac{\Gamma}{\Np}}}
\newcommand{\etaIth}{{\eta_{I\mathrm{th}}}}
\newcommand{\etaI}{{\eta_I}}
\newcommand{\nst}{{n^\mathrm{st}_\mathrm{ph}}}
\newcommand{\nsp}{{n^\mathrm{sp}_\mathrm{ph}}}
\newcommand{\nph}{{n_\mathrm{ph}}}
\newcommand{\tpara}{{\tau_{3,\mathrm{para}}}}
\newcommand{\tsp}{{\tau_\mathrm{sp}}}
\newcommand{\tph}{{\tau_\mathrm{ph}}}
\newcommand{\tpsp}{{\tau'_\mathrm{sp}}}
\newcommand{\tst}{{\tau_\mathrm{st}}}
\newcommand{\gmat}{{g_{\mathrm{mat}}}}
\newcommand{\Ith}{{I_{\mathrm{th}}}}
\newcommand{\Ithp}{{I_{\mathrm{th}}^+}}
\newcommand{\Ithm}{{I_{\mathrm{th}}^-}}
\newcommand{\Jth}{{J_{\mathrm{th}}}}
\newcommand{\Jtht}{{J_{\mathrm{th},10 \mathrm{K}}}}
\newcommand{\Jthf}{{J_{\mathrm{th},5 \mathrm{K}}}}
\newcommand{\Jmaxf}{{J_{\mathrm{max},5 \mathrm{K}}}}
\newcommand{\Tactive}{{T_{\mathrm{active}}}}
\newcommand{\Tcw}{{T_{\mathrm{cw}}}}
\newcommand{\Tmax}{{T_{\mathrm{max}}}}
\newcommand{\Tmaxp}{{T_{\mathrm{max,pul}}}}
\newcommand{\Tmaxc}{{T_{\mathrm{max,cw}}}}
\newcommand{\Jmax}{{J_{\mathrm{max}}}}
\newcommand{\Jpeak}{{J_{\mathrm{peak}}}}
\newcommand{\Imax}{{I_{\mathrm{max}}}}
\newcommand{\Vac}{{V_{\mathrm{ac}}}}
\newcommand{\Vmod}{{V_{\mathrm{mod}}}}
\newcommand{\Aac}{{A_{\mathrm{ac}}}}
\newcommand{\Vcav}{{V_{\mathrm{cav}}}}
\newcommand{\Acav}{{A_{\mathrm{cav}}}}
\newcommand{\SEQUAL}{\mbox{SEQUAL}}
\newcommand{\mez}{{m^{\ast}(z)}}
\newcommand{\me}{m^\ast}
\newcommand{\e}{m^{\ast}(z,E)}
\newcommand{\mezei}{m^{\ast}(z,E_i)}
\newcommand{\mezef}{m^{\ast}(z,E_f)}
\newcommand{\ddz}{\frac{\partial}{\partial z}}
\newcommand{\pddt}{\frac{\partial}{\partial t}}
\newcommand{\ddt}{\frac{d}{d t}}
\newcommand{\etal}{{\it {et~al.\ }}}
\newcommand{\ie}{\mbox{i.\ e.}}
\newcommand{\psicz}{\psi_c(z)}
\newcommand{\psiciz}{\psi_c^{(i)}(z)}
\newcommand{\psicfz}{\psi_c^{(f)}(z)}
\newcommand{\epsc}{\epsilon_{\mathrm{core}}}
\newcommand{\epsmc}{\epsilon_{\mathrm{m,core}}}
\newcommand{\epsr}{\epsilon_{\mathrm{r}}}
\newcommand{\epsm}{\epsilon_{\mathrm{m}}}
\newcommand{\epsd}{\epsilon_{\mathrm{d}}}
\newcommand{\epsdt}{\epsilon^2_{\mathrm{d}}}
\newcommand{\epsz}{\epsilon_{0}}
\newcommand{\km}{k_\mathrm{m}}
\newcommand{\kd}{k_\mathrm{d}}
\newcommand{\kdt}{k^2_\mathrm{d}}
\newcommand{\degC}[1]{{#1\,^\circ\mathrm{C}}}
\newcommand{\betaz}{{\beta_z}}
\newcommand{\neff}{{n_\mathrm{eff}}}
\newcommand{\wpl}{{\omega_{\mathrm{p}}}}
\newcommand{\wplt}{{\omega^2_{\mathrm{p}}}}
\newcommand{\wspl}{{\omega_{\mathrm{sp}}}}
\newcommand{\kpari}{{\mathbf{k}_{\|,i}}}
\newcommand{\kparf}{{\mathbf{k}_{\|,f}}}
\newcommand{\kparm}{{\mathbf{k}_{\|,m}}}
\newcommand{\kparn}{{\mathbf{k}_{\|,n}}}
\newcommand{\kpara}{{\mathbf{k}_{\|,a}}}
\newcommand{\kparb}{{\mathbf{k}_{\|,b}}}
\newcommand{\ikpari}{{i,{\mathbf k}_{\|,i}}}
\newcommand{\fkparf}{{f,{\mathbf k}_{\|,f}}}
\newcommand{\mkparm}{{m,{\mathbf k}_{\|,m}}}
\newcommand{\akpara}{{a,{\mathbf k}_{\|,a}}}
\newcommand{\bkparb}{{b,{\mathbf k}_{\|,b}}}
\newcommand{\phinr}{{\phi_n \rangle}}
\newcommand{\phiml}{{\langle \phi_m}}
\newcommand{\psitket}{{|\psi(t)\rangle}}
\newcommand{\iket}{{| i \rangle}}
\newcommand{\jket}{{| j \rangle}}
\newcommand{\aket}{{| a \rangle}}
\newcommand{\bket}{{| b \rangle}}
\newcommand{\ket}[1]{{| #1 \rangle}}
\newcommand{\bra}[1]{{\langle #1 |}}
\newcommand{\ibra}{{\langle i |}}
\newcommand{\jbra}{{\langle j |}}
\newcommand{\abra}{{\langle a |}}
\newcommand{\bbra}{{\langle b |}}
\newcommand{\kpar}{{\mathbf{k}_{\|}}}
\newcommand{\ahat}{{\hat{a}}}
\newcommand{\Vbias}{{V_\mathrm{bias}}}
\newcommand{\ahatc}{{\hat{a}^\dagger}}
\newcommand{\eikr}{{e^{i \mathbf{k}\cdot\mathbf{r}}}}
\newcommand{\emikr}{{e^{-i \mathbf{k}\cdot\mathbf{r}}}}
\newcommand{\br}{{\mathbf{r}}}
\newcommand{\bz}{{\mathbf{z}}}
\newcommand{\bk}{{\mathbf{k}}}
\newcommand{\bu}{{\mathbf{u}}}
\newcommand{\usigma}{{\mathbf{u}_\sigma}}
\newcommand{\ksigma}{{\mathbf{k},\sigma}}
\newcommand{\rpar}{{\mathbf{r}_\|}}
\newcommand{\bA}{{\mathbf A}}
\newcommand{\bE}{{\mathbf E}}
\newcommand{\ki}{{\mathbf{k}_i}}
\newcommand{\kf}{{\mathbf{k}_f}}
\newcommand{\kj}{{\mathbf{k}_j}}
\newcommand{\kg}{{\mathbf{k}_g}}
\newcommand{\qpara}{{\bf q}_{\parallel}}
\newcommand{\qz}{{\bf q}_z}
\newcommand{\LO}{{\mathrm{LO}}}
\newcommand{\hwLO}{{\hbar\omega_{\mathrm{LO}}}}
\newcommand{\wLO}{{\omega_{\mathrm{LO}}}}
\newcommand{\ELO}{{E_{\mathrm{LO}}}}
\newcommand{\bracomboa}{{| n_a^\ksigma : \akpara \rangle}}
\newcommand{\bracombob}{{| n_b^\ksigma : \bkparb \rangle}}
\newcommand{\comboa}{{n_a^\ksigma : \akpara}}
\newcommand{\combob}{{n_b^\ksigma : \bkparb}}
\newcommand{\mat}{{\mathrm{mat}}}
\newcommand{\rmmin}{{\mathrm{min}}}
\newcommand{\rmmax}{{\mathrm{max}}}
\newcommand{\rmpeak}{{\mathrm{peak}}}
\newcommand{\rmph}{{\mathrm{ph}}}
\newcommand{\rmps}{{\mathrm{ps}}}
\newcommand{\rmst}{{\mathrm{st}}}
\newcommand{\rmsp}{{\mathrm{sp}}}
\newcommand{\thz}{{\mathrm{THz}}}
\newcommand{\meV}{{\mathrm{meV}}}
\newcommand{\eV}{{\mathrm{eV}}}
\newcommand{\ps}{{\mathrm{ps}}}
\newcommand{\rma}{{\mathrm{a}}}
\newcommand{\rms}{{\mathrm{s}}}
\newcommand{\rmand}{{\mathrm{and}}}
\newcommand{\rmwhere}{{\mathrm{where}}}
\newcommand{\rmpara}{{\mathrm{para}}}
\newcommand{\cav}{{\mathrm{cav}}}
\newcommand{\thD}{{\mathrm{3D}}}
\newcommand{\twD}{{\mathrm{2D}}}
\newcommand{\rmeff}{{\mathrm{eff}}}
\newcommand{\rmhot}{{\mathrm{hot}}}
\newcommand{\rmLO}{{\mathrm{LO}}}
\newcommand{\rmmW}{{\mathrm{mW}}}
\newcommand{\rmuW}{{\mu\mathrm{W}}}
\newcommand{\rmK}{{\mathrm{K}}}
\newcommand{\nGaAs}{{n_\mathrm{GaAs}}}
\newcommand{\nSi}{{n_\mathrm{Si}}}
\newcommand{\rmGaAs}{{\mathrm{GaAs}}}
\newcommand{\slopeeff}{{\frac{\alpha_{\rmm,\rmf}}{\alpha_\rmw+\alpha_{\rmm,\rmf}+\alpha_{\rmm,\rmr}}}}
\newcommand{\rmfor}{{\mathrm{for}}}
\newcommand{\inlimit}{{\mathrm{in \ the \ limit}}}
\newcommand{\rmtb}{{\mathrm{tb}}}
\newcommand{\rmr}{{\mathrm{r}}}
\newcommand{\rmf}{{\mathrm{f}}}
\newcommand{\rme}{{\mathrm{e}}}
\newcommand{\rmR}{{\mathrm{R}}}
\newcommand{\rmi}{{\mathrm{i}}}
\newcommand{\rmnm}{{\mathrm{nm}}}
\newcommand{\rmu}{{\mathrm{u}}}
\newcommand{\rml}{{\mathrm{l}}}
\newcommand{\rmj}{{\mathrm{j}}}
\newcommand{\rmm}{{\mathrm{m}}}
\newcommand{\rmw}{{\mathrm{w}}}
\newcommand{\rmmm}{{\mathrm{mm}}}
\newcommand{\rmmat}{{\mathrm{mat}}}
\newcommand{\rmth}{{\mathrm{th}}}
\newcommand{\rmV}{{\mathrm{V}}}
\newcommand{\defas}{{\stackrel{\triangle}{=}}}
\newcommand{\itof}{{i \rightarrow f}}
\newcommand{\ftoi}{{f \rightarrow i}}
\newcommand{\atob}{{a \rightarrow b}}
\newcommand{\btoa}{{b \rightarrow a}}
\newcommand{\qsig}{{{\bf q},\sigma}}
\newcommand{\epol}{{\bf \hat{e}}_{{\bf q},\sigma}}
\newcommand{\zhat}{{\bf \hat{z}}}
\newcommand{\xhat}{{\bf \hat{x}}}
\newcommand{\pop}{{\hat{\mathbf{p}}}}
\newcommand{\rop}{{\hat{\mathbf{r}}}}
\newcommand{\rzop}{{\hat{\mathbf{z}}}}
\newcommand{\rpop}{{\hat{\mathbf{r}}_{\|}}}
\newcommand{\Aop}{{\hat{\mathbf{A}}}}
\newcommand{\Eop}{{\hat{\mathbf{E}}}}
\newcommand{\Hop}{{\hat{H}}}
\newcommand{\Oop}{{\hat{O}}}
\newcommand{\Omat}{{\bar{\bar{O}}}}
\newcommand{\Htbop}{{\hat{H}_\mathrm{tb}}}
\newcommand{\Htbmat}{{\bar{\bar{H}}_\mathrm{tb}}}
\newcommand{\Hext}{{\hat{H}}}
\newcommand{\Hextop}{{\hat{H}_\mathrm{ext}}}
\newcommand{\Hextmat}{{\bar{\bar{H}}_\mathrm{ext}}}
\newcommand{\Hmat}{{\bar{\bar{H}}}}
\newcommand{\rhomat}{{\bar{\bar{\rho}}}}
\newcommand{\rhoop}{{\hat{\rho}}}
\newcommand{\ntot}{{n_\mathrm{tot}}}
\newcommand{\Lmod}{{L_\mathrm{mod}}}
\newcommand{\Lcav}{{L_\mathrm{cav}}}
\newcommand{\tulhotLO}{{\tau^\rmhot_{ul,\rmLO}}}
\newcommand{\tulLO}{{\tau_{ul,\rmLO}}}
\newcommand{\tFLparb}{{\tau_{5\rightarrow(6,3,2,1)}}}
\newcommand{\tFLpar}{{\tau_{5,\mathrm{par}}}}
\newcommand{\Tpure}{{T^*_2}}
\newcommand{\td}{{\tau_{\|}}}
\newcommand{\tdgen}[1]{{\tau_{\|#1}}}
\newcommand{\tdtw}{{\tau_{\|,2}}}
\newcommand{\tdtwp}{{\tau_{\|,2'}}}
\newcommand{\tdtwpsq}{{\tau^2_{\|,2'}}}
\newcommand{\tdtwth}{{\tau_{\|,23}}}
\newcommand{\tdtwpth}{{\tau_{\|,2'3}}}
\newcommand{\tdth}{{\tau_{\|,3}}}
\newcommand{\tdsq}{{\tau^2_{\|}}}
\newcommand{\Omsqgen}[1]{{\Omega_{#1}^2}}
\newcommand{\Delsqgen}[1]{{\Delta_{#1}^2}}
\newcommand{\tdsqgen}[1]{{\tau^2_{\|#1}}}
\newcommand{\Ahsqgen}[1]{{\left(\frac{\Delta_{#1}}{\hbar}\right)^2}}
\newcommand{\Ahsqgenb}[1]{{\Omega_{#1}^2}}
\newcommand{\Ahsq}{{\left(\frac{\Delta_0}{\hbar}\right)^2}}
\newcommand{\Ahcsq}{{\left(\frac{\delzc}{\hbar}\right)^2}}
\newcommand{\Ehsqgen}[1]{{\left(\frac{E_{#1}}{\hbar}\right)^2}}
\newcommand{\Ehsq}{{\left(\frac{E_{1'3}}{\hbar}\right)^2}}
\newcommand{\Ehcsq}{{\left(\frac{E_{22'}}{\hbar}\right)^2}}
\newcommand{\delop}{{\hat{\nabla}}}
\newcommand{\zzif}{|z_{i \rightarrow f}|^2}
\newcommand{\deltakr}{{\delta^{\mathrm{kr}}}}
\newcommand{\delnu}{{\Delta \nu}}
\newcommand{\delfwhm}{{\Delta \nu_\mathrm{FWHM}}}
\newcommand{\Erad}{{E_\mathrm{rad}}}
\newcommand{\frad}{{f_\mathrm{rad}}}
\newcommand{\zrad}{{z_\mathrm{rad}}}
\newcommand{\nurad}{{\nu_\mathrm{rad}}}
\newcommand{\delec}{{\Delta E_\mathrm{c}}}
\newcommand{\delnth}{{\Delta n_\mathrm{th}}}
\newcommand{\drr}{{\Delta \mathcal{R}_\mathrm{th}/\mathcal{R}_\mathrm{th}}}
\newcommand{\drrfrac}{{\frac{\Delta \mathcal{R}_\mathrm{th}}{\mathcal{R}_\mathrm{th}}}}
\newcommand{\drrfracb}{{{\Delta \mathcal{R}_\mathrm{th}}/{\mathcal{R}_\mathrm{th}}}}
\newcommand{\Rth}{{\mathcal{R}_\mathrm{th}}}
\newcommand{\delzc}{{\Delta^{\mathrm{c}}_0}}
\newcommand{\gth}{{g_\mathrm{th}}}
\newcommand{\delN}{{\Delta N}}
\newcommand{\icm}{{{\mathrm{cm}}^{-1}}}
\newcommand{\Appcm}{{{\mathrm{A}/\mathrm{cm}}^{2}}}
\newcommand{\iicm}{{{\mathrm{cm}}^{-2}}}
\newcommand{\iiicm}{{{\mathrm{cm}}^{-3}}}
\newcommand{\iiium}{{{\mu\mathrm{m}}^{-3}}}
\newcommand{\iiim}{{{\mathrm{m}}^{-3}}}
\newcommand{\kacmm}{{\rm kA/cm}^2}
\newcommand{\real}[1]{{\mathcal{R}\mathrm{e}\{#1\}}}
\newcommand{\imag}[1]{{\mathcal{I}\mathrm{m}\{#1\}}}
\newcommand{\um}{{\mu\mathrm{m}}}
\newcommand{\IV}{{$I$-$V$}}
\newcommand{\LIV}{{$L$-$I$-$V$}}
\newcommand{\VIs}{{$V$-$I$s}}
\newcommand{\IVs}{{$I$-$V$s}}
\newcommand{\VI}{{$V$-$I$}}
\newcommand{\LI}{{$L$-$I$}}
\newcommand{\LIs}{{$L$-$I$s}}
\newcommand{\LVs}{{$L$-$V$s}}
\newcommand{\PI}{{$P$-$I$}}
\newcommand{\LV}{{$L$-$V$}}
\newcommand{\RV}{{$\mathcal{R}$-$V$}}
\newcommand{\RVs}{{$\mathcal{R}$-$V$s}}
\newcommand{\RI}{{$\mathcal{R}$-$I$}}
\newcommand{\RIs}{{$\mathcal{R}$-$I$s}}
\newcommand{\JV}{{$J$-$V$}}
\newcommand{\GV}{{$G$-$V$}}
\newcommand{\CV}{{$C$-$V$}}
\newcommand{\IB}{{$I$-$B$}}
\newcommand{\GB}{{$G$-$B$}}
\newcommand{\dVdI}{{$dV/dI$}}
\newcommand{\dVdIV}{{$dV/dI$-$V$}}
\newcommand{\dVdII}{{$dV/dI$-$I$}}
\newcommand{\ebar}{\overline{\eta}}
\newcommand{\Perot}{{P\'{e}rot}}
\newcommand{\phos}{{$\mathrm{H_3PO_4:H_2O_2:H_2O}$}}
\newcommand{\sulf}{{$\mathrm{H_2SO_4:H_2O_2:H_2O}$}}
\newcommand{\amm}{{$\mathrm{NH_4OH:H_2O_2:H_2O}$}}
\newcommand{\ammlap}{{$\mathrm{NH_4OH:H_2O_2}$}}
\newcommand{\alumina}{{$\mathrm{Al_2O_3}$}}
\newcommand{\alcoat}{{$\mathrm{Al_2O_3/Ti/Au/Al_2O_3}$}}
\newcommand{\ang}{{\mathrm{\AA}}}

\preprint{}

\title{Terahertz plasmonic laser radiating in an ultra-narrow beam}

\author{Chongzhao Wu,$^1$ Sudeep Khanal,$^1$ John L. Reno,$^2$ and Sushil Kumar$^{1}$}

\affiliation{$^1$Department of Electrical and Computer Engineering, Lehigh University, Bethlehem, PA 18015\\}

\affiliation{$^2$Sandia National Laboratories, Center of Integrated Nanotechnologies, MS 1303, Albuquerque, NM 87185-1303\\}

\email{chw310@lehigh.edu,sushil@lehigh.edu}

\date{\today}

\begin{abstract}
Plasmonic lasers (spasers) generate coherent surface-plasmon-polaritons
(SPPs) and could be realized at subwavelength dimensions in metallic cavities
for applications in nanoscale optics.
Plasmonic cavities are also utilized for terahertz quantum-cascade
lasers (QCLs), which are the brightest available solid-state
sources of terahertz radiation. A long standing challenge for spasers that are utilized as nanoscale sources of radiation, is their poor coupling to the far-field radiation. Unlike conventional lasers that could produce
directional beams, spasers have highly divergent radiation patterns due to their subwavelength
apertures. Here, we theoretically and experimentally demonstrate a new technique
for implementing distributed-feedback (DFB) 
that is distinct from any other previously utilized DFB schemes for semiconductor lasers.
The so-termed {\em antenna-feedback} scheme leads to single-mode operation in plasmonic lasers,
couples the resonant SPP mode to a highly directional far-field radiation pattern, and
integrates hybrid SPPs in surrounding medium into the operation of the DFB lasers.
Experimentally, the antenna-feedback method, which does not require the phase matching to a well-defined effective index, is implemented for terahertz QCLs, and 
single-mode terahertz QCLs with beam divergence as small as $4^\circ\times 4^\circ$ are demonstrated,
which is the narrowest beam reported for any terahertz QCL to-date. Moreover, in contrast to
negligible radiative-field in conventional photonic band-edge lasers, in which the periodicity follows the integer multiple of half-wavelength inside active medium,   antenna-feedback breaks this integer-limit for the first time and
enhances the radiative-field of lasing mode.
Terahertz lasers with narrow-beam emission will find applications for integrated as well as standoff terahertz
spectroscopy and sensing. The antenna-feedback scheme is generally applicable to any plasmonic laser with a \FPerot\ cavity
irrespective of its operating wavelength, and could bring plasmonic lasers closer to practical applications.
\end{abstract}


\maketitle
\thispagestyle{fancy}

\section{Introduction}
\label{introduction}

A surface-plasmon-polariton (SPP) is a coupled state between electromagnetic (EM) field and electron plasma
oscillations at the interface between a metal and a dielectric, for which the EM field could be confined in
subwavelength dimensions normal to the surface of metal. Consequently, metallic cavities
supporting SPP modes have been used to realize SPP lasers (also known as plasmonic lasers or spasers) with
subwavelength dimensions~\cite{bergman:spaser,hill:smalllasers,oulton:splasers,berini:spplasers,noginov:spaser}.
The energy in a spaser can remain confined as coherent SPPs or it can be made to leak out from the spaser as radiation. In many targeted applications in integrated optics and nanophotonics, spasers are developed as nanoscale sources of coherent electromagnetic (EM) radiation and show interesting properties such as ultrafast dynamics for applications in high-speed optical communications. Parallel-plate metallic
cavities supporting SPP modes are also utilized for terahertz quantum-cascade lasers
(QCLs)~\cite{kohler:laser} to achieve low-threshold and high-temperature performance~\cite{williams:review}
owing to the low-loss of SPP modes at terahertz frequencies that are much smaller than
the plasma frequency in metal. The most common type of plasmonic lasers with long-range SPPs, which include
terahertz QCLs, utilize \FPerot\ type cavities in which at least one dimension is longer
than the wavelength inside the dielectric~\cite{hill:mimlaser,oulton:plasmonlaser,lu:nanolaser,williams:metal}.
One of the most important challenges for such plasmonic lasers is their poor coupling to free-space radiation
modes owing to the subwavelength mode confinement in the cavity, which leads to small radiative efficiency
and highly divergent radiation patterns. This problem is also severe for terahertz QCLs based on metallic
cavities and leads to low-output power and undesirable omnidirectional radiation patterns from
\FPerot\ cavities~\cite{adam:beam,orlova:wirelaser}.

A possible solution to achieve directionality
of far-field emission from spasers is to utilize periodic structures with broad-area emission, which has
been used for both short-wavelength spasers~\cite{zhou:nanocavityarrays, beijnum:holearrays,
meng:spaserarray,yang:real,schokker:lasing,dorofeenko:steady} as well
as terahertz QCLs~\cite{mahler:review11,sirtori:review13}. On chip phased-locked arrays~\cite{kao:phase,halioua:phaselocked} or metasurface reflectors composed of multiple cavities~\cite{xu:metasurface} have also been utilized for directional emission in terahertz QCLs. However, edge-emitting \FPerot\ cavity structures with narrow cavity widths are more desirable, especially for electrically pumped spasers to achieve small operating electrical power and better heat removal from the cavity (along the width of the cavity in lateral dimension, through the substrate) for continuous-wave (cw) operation. In this paper, we theoretically and experimentally demonstrate
a new technique for implementing distributed-feedback (DFB) in plasmonic lasers with \FPerot\ cavities, which
is termed as an {\em antenna-feedback} scheme. This DFB scheme has no resemblance to the 
the multitude of DFB methods that have been conventionally utilized for semiconductor lasers.
The key concept is to couple the guided SPP mode in a spaser’s cavity to a single-sided SPP mode that can
exist in its surrounding medium by periodic perturbation of the metallic cladding in the cavity. 
Such a coupling is possible by choosing a Bragg grating of appropriate periodicity in the metallic film.
This leads to excitation of coherent single-sided SPPs on the metallic cladding of the spaser, which couple to a narrow-beam
in the far-field. 
The narrow beam emission is due in part to the cavity acting like an end-fire phased-array antenna at the
microwave frequencies, as well as due to the large spatial extent of a coherent single-sided SPP mode that
is generated on the metal film as a result of the feedback scheme. Experimentally, the antenna-feedback
method is implemented for terahertz QCLs for which the method is shown to be an improvement
over the recently developed third-order DFB scheme for producing directional beams~\cite{amanti:dfb}
since it does not require any specific design considerations for phase-matching~\cite{kao:dfbpm}. The emitted
beam is more directional and the output power is also increased due to increased radiative field by
virtue of this specific scheme. 
\begin{figure*}[htbp]
\centering
\includegraphics[width=6.5in]{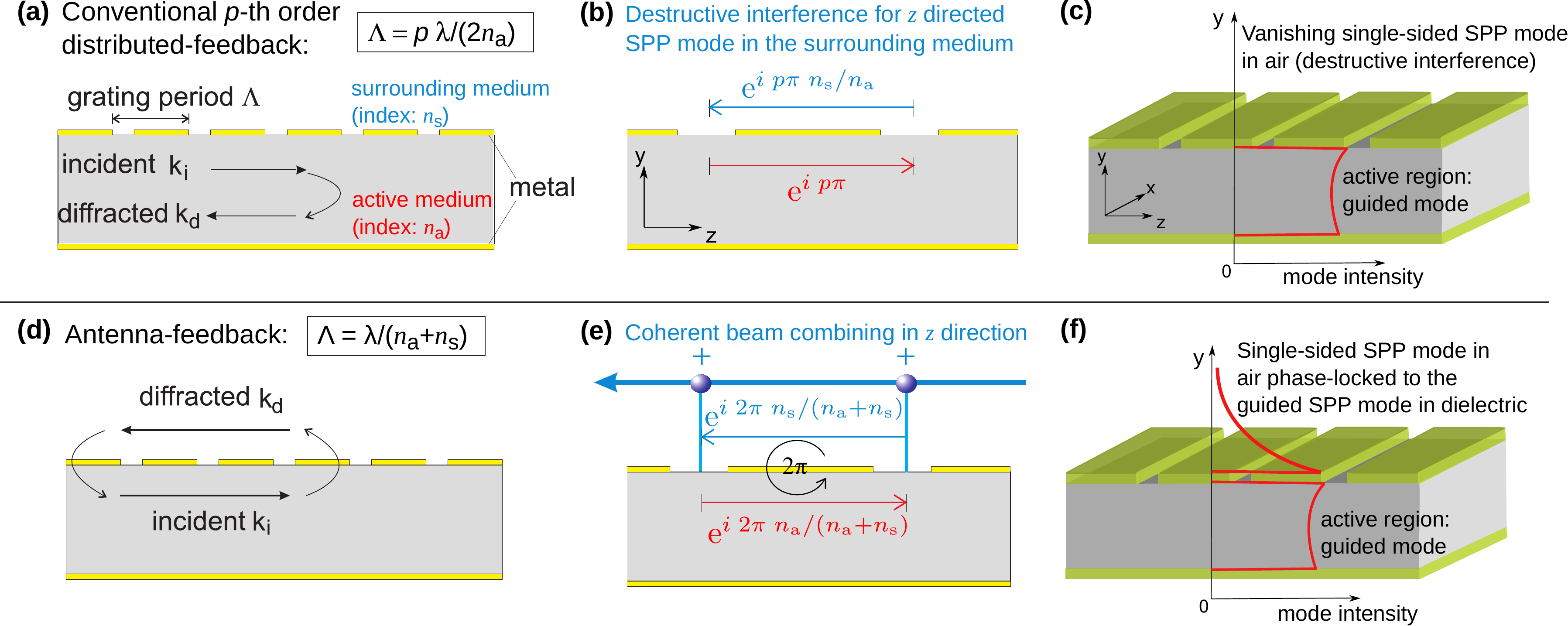}
\vspace{-0.10in}
\caption{
{\bf The antenna-feedback concept for spasers.} 
(a)~The general principle of conventional distributed-feedback (DFB) that could be
implemented in a spaser by introducing periodicity in its metallic cladding.
A parallel-plate metallic cavity is illustrated; however, the principle is equally applicable to
spaser cavities with a single metal-cladding.
(b) If the periodicity in (a) is implemented by making holes or slits in the metal-cladding,  
the guided SPP wave diffracts out through the apertures and generates single-sided SPP waves on 
the cladding in the surrounding medium. The figure shows phase-mismatch between successive apertures
for SPP waves on either side of the cladding. Coherent single-sided SPP waves in the surrounding medium cannot therefore
be sustained owing to destructive interference with the guided SPP wave inside the cavity, as illustrated in~(c).
(d)~Principle of an antenna-feedback grating. If the periodicity in the metal film allows the guided
SPP mode to diffract outside the cavity, a grating period could
be chosen that leads to the first-order Bragg diffraction in the opposite direction, but in
the surrounding medium rather than inside the active medium itself.
Similarly, the single-sided SPP mode in the surrounding medium undergoes first-order Bragg diffraction to couple with the guided SPP
wave in the opposite direction inside the cavity.
(e) The grating in (d) leads to a fixed phase-condition at each aperture between counter propagating SPP waves on the
either side of metal-cladding. First, this leads to significant build up of amplitude
in the single-sided SPP wave in the surrounding medium, as illustrated in (f). Second,
emission from each aperture adds constructively to couple to far-field radiation in the end-fire ($z$) direction.
As argued in the text, both of these aspects lead to narrow far-field emission profile in the $x-y$ plane.
\vspace{-0.1in}
}
\label{fig1}
\end{figure*}

\section{Antenna-feedback scheme for plasmonic lasers}
\label{results}

Single-mode operation in spasers with \FPerot\ cavities could be implemented in a straightforward manner 
by periodically perturbing the metallic film that supports the resonant SPP modes. The schematic in
Fig.~\ref{fig1}(a) shows an example of a periodic grating in the top metal cladding for a parallel-plate
metallic cavity that could be utilized to implement conventional $p$-th order DFB by choosing the appropriate periodicity.
Since the SPP mode has maximum amplitude at the interface of metal and dielectric active medium, a periodic
perturbation in the metal film could provide strong Bragg diffraction up to high-orders for the counter-propagating
SPP waves inside the active medium with incident and diffracted wavevectors $k_i$ and $k_d=-k_i$ respectively,
such that
\begin{eqnarray}
k_i & = & p \frac{2\pi}{\Lambda} + k_d
\label{eq1}
\end{eqnarray}
where $\Lambda$ is the grating period, $2\pi/\Lambda$ is the grating wavevector, and $p$ is an integer
($p=1,2,3\ldots$) that specifies the diffraction order.
For plane-wave like modes at frequencies far away from the plasma resonance in metal, $k_i\approx 2\pi n_\rma/\lambda$,
where $\lambda$ is the free-space wavelength corresponding to the SPP mode and $n_\rma$ is
the effective propagation index in active medium (approximately the same as refractive index of the medium), the so-called
Bragg mode with $\lambda=2 n_\rma \Lambda/p$ is resonantly
excited because it is, by design, the lowest-loss mode in the DFB cavity within the gain spectrum of the active medium.
For terahertz QCLs with metallic cavities, first-order~\cite{williams:corrdfb} and
second-order~\cite{fan:surfemit,kumar:surfemit} DFBs have been implemented to achieve
robust single-mode operation. However, these conventional DFB techniques do not achieve directionality of
far-field radiation in both directions. There is phase mismatch for SPP waves on either side of metal claddings and destructive interference between successive apertures, as shown in Fig.~\ref{fig1}(b) for propagating SPP waves. Therefore, no coherent single-sided SPP waves can be established on the metallic cladding in the surrounding medium as demonstrated in Fig.~\ref{fig1}(c). Consequently, 2D photonic-crystal DFB structures have been utilized for
broad-area (surface) single-mode emission~\cite{chassagneux:nature,halioua:phocrystal} for which
diffraction-limited beams could be achieved at the expense of large cavity dimensions.

In contrast to conventional DFB methods in which periodic gratings couple
forward and backward propagating waves inside the active medium itself, the
antenna-feedback scheme couples a single-sided SPP wave that travels in the surrounding medium
with the SPP wave traveling inside the active medium as illustrated in Fig.~\ref{fig1}(d).
The SPP wave inside the active medium with incident wavevector $k_i\approx 2\pi n_\rma/\lambda$
is diffracted in the opposite direction in the surrounding medium with wavevector $k_d\approx -2\pi n_\rms/\lambda$.
For the first-order diffraction grating ($p=1$), equation~\eqref{eq1} results in
\begin{eqnarray}
\frac{2\pi n_\rma}{\lambda} & = & \frac{2\pi}{\Lambda} - \frac{2\pi n_\rms}{\lambda}
\label{eq2}
\end{eqnarray}
that leads to excitation of a DFB mode with $\lambda=(n_\rma+n_\rms) \Lambda$, which is
different from any of the $p$-th order DFB modes that occur at $\lambda=2 n_\rma \Lambda/p$. Hence,
the antenna-feedback mode could always be excited by just selecting the appropriate grating period
$\Lambda$ such that the wavelength occurs close to the peak-gain wavelength in the active-medium.  
For GaAs/AlGaAs based terahertz QCLs, $n_\rma\sim 3.6$, $n_\rms=1$, and hence for a chosen grating-period
$\Lambda$, the first-order DFB, antenna-feedback, and second-order DFB modes occur at wavelengths
$7.2~\Lambda$, $4.6~\Lambda$, and $3.6~\Lambda$ respectively. The typical gain-bandwidth
of terahertz QCLs is less than $20\%$ of the peak-gain wavelength, which suggests that the grating
has to be designed specifically to excite the antenna-feedback mode.

\begin{figure*}[htbp]
\centering
\includegraphics[width=6.5in]{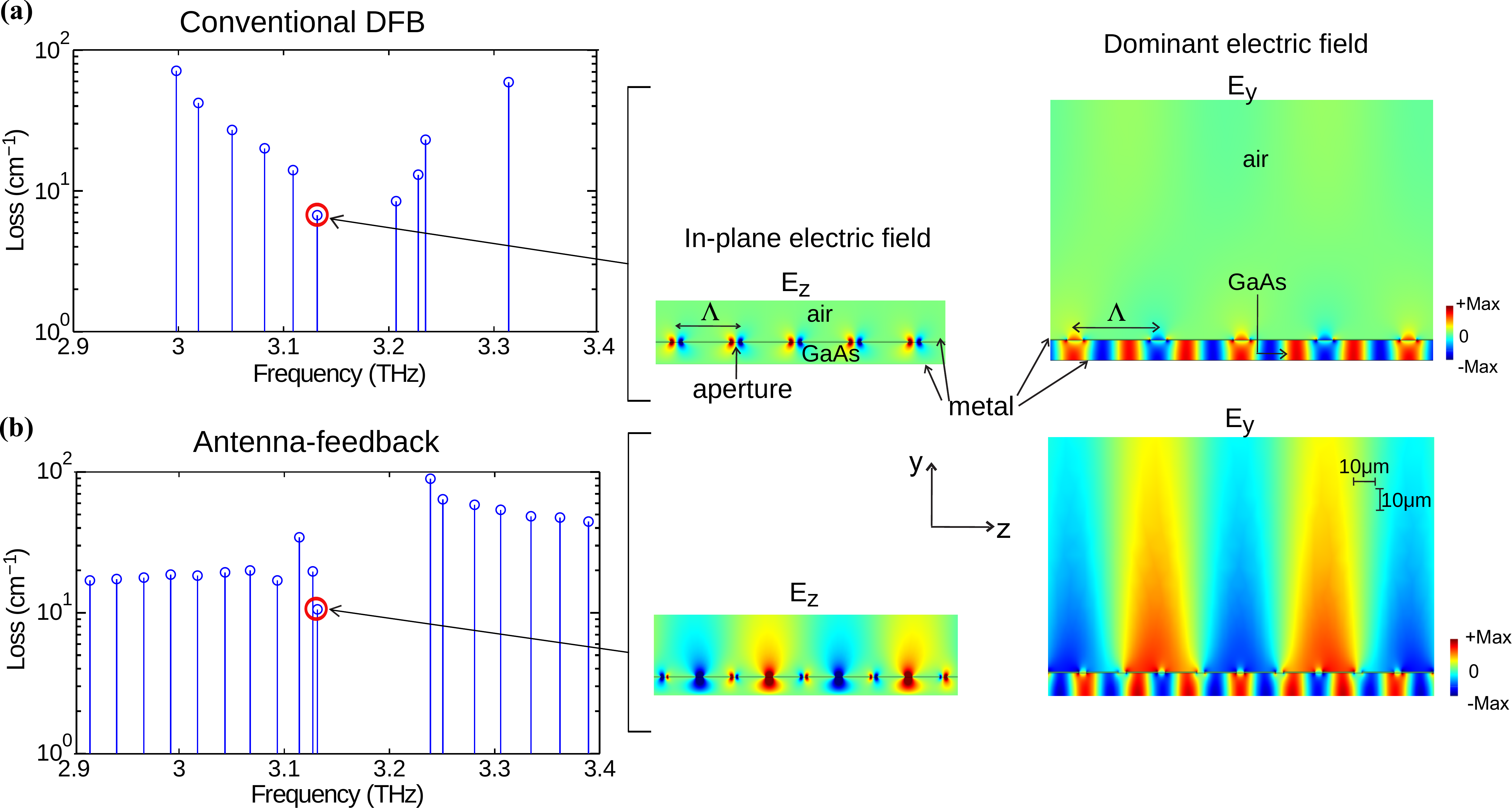}
\caption{
{\bf Comparison between conventional DFB (third-order DFB as an example) and antenna-feedback schemes for terahertz QCL cavities.}
The figure shows SPP eigenmode spectrum and electric-field for the eigenmode with lowest loss calculated by
finite-element simulations of parallel-plate metallic cavities as in Fig.~\ref{fig1}, with GaAs as dielectric
($n_\rma=3.6$) and air as surrounding medium ($n_\rms=1$). Simulations are done in 2-D (\ie\ cavities of 
infinite-width) for $10~\um$ thick and $1.4~$mm long cavities, and metal and active-layers are considered lossless.
Lossy sections are implemented in the cavities
at both longitudinal ends of the cavity as absorbing boundaries, which eliminates
reflection of guided SPP modes from the end-facets.
A periodic grating with apertures of (somewhat arbitrary) width
$0.2\Lambda$ in the top-metal cladding are implemented for DFB. $\Lambda$ is chosen to excite the
lowest-loss DFB mode at similar frequencies close to $\sim 3~\thz$. The eigenmode spectrum shows frequencies and
loss for the resonant-cavity modes, which reflects combination of radiation loss and the loss at
longitudinal absorbing regions. 
Fig.~(a) shows results from a third-order DFB grating with $\Lambda=41.7~\um$ and (b) shows results from
antenna-feedback grating with $\Lambda=21.7~\um$. Radiation loss occurs through diffraction from
apertures, and the amplitude of in-plane electric-field $E_z$ is indicative of the outcoupling efficiency.
The major fraction of EM energy for the resonant modes exists in TM polarized ($E_y$) electric-field.
A photonic bandgap in the eigenmode spectrum is indicative of DFB effect due to the grating.
The antenna-feedback grating excites a strong single-sided SPP standing-wave on top of the metallic grating (in air)
as also illustrated in Fig.~\ref{fig1}(f). Also, the radiative loss for the third-order DFB grating is smaller
since the lowest-loss eigenmode has zeros of $E_z$ under the apertures, which leads to smaller
net outcoupling of radiation. The loss is $\sim 6.7~\icm$ and $\sim 10.6~\icm$ for the lowest loss resonant cavity mode of third-order DFB  and antenna-feedback scheme, respectively.  
}
\label{fig2}
\end{figure*}

The antenna-feedback scheme leads to excitation of a coherent
single-sided SPP standing-wave on the metallic cladding of the spaser, which is phase-locked to
the resonant-cavity SPP mode inside the active medium as shown in Fig.~\ref{fig1}(f)..
Both waves maintain exact same phase relation at each aperture location,
where they exchange electromagnetic (EM) energy with each other due to diffraction
as illustrated in Fig.~\ref{fig1}(e). The SPP wave in the surrounding medium is excited due to 
scattering of EM field at apertures that generates a combination of propagating quasi-cylindrical
waves and SPPs~\cite{lalanne:creepingwave2,babuty:mirspsource}
that propagate along the surface of the metal-film. The scattered waves thus generated at each
aperture superimpose constructively in only the end-fire ($z$) direction owing to the phase-condition
thus established at each aperture. For coupling to far-field radiation, the radiation is therefore analogous
to that from an end-fire phased array antenna that produces a narrow beam in both $x$ and $y$ directions.

\begin{figure}[htbp]
\centering
\includegraphics[width=3.3in]{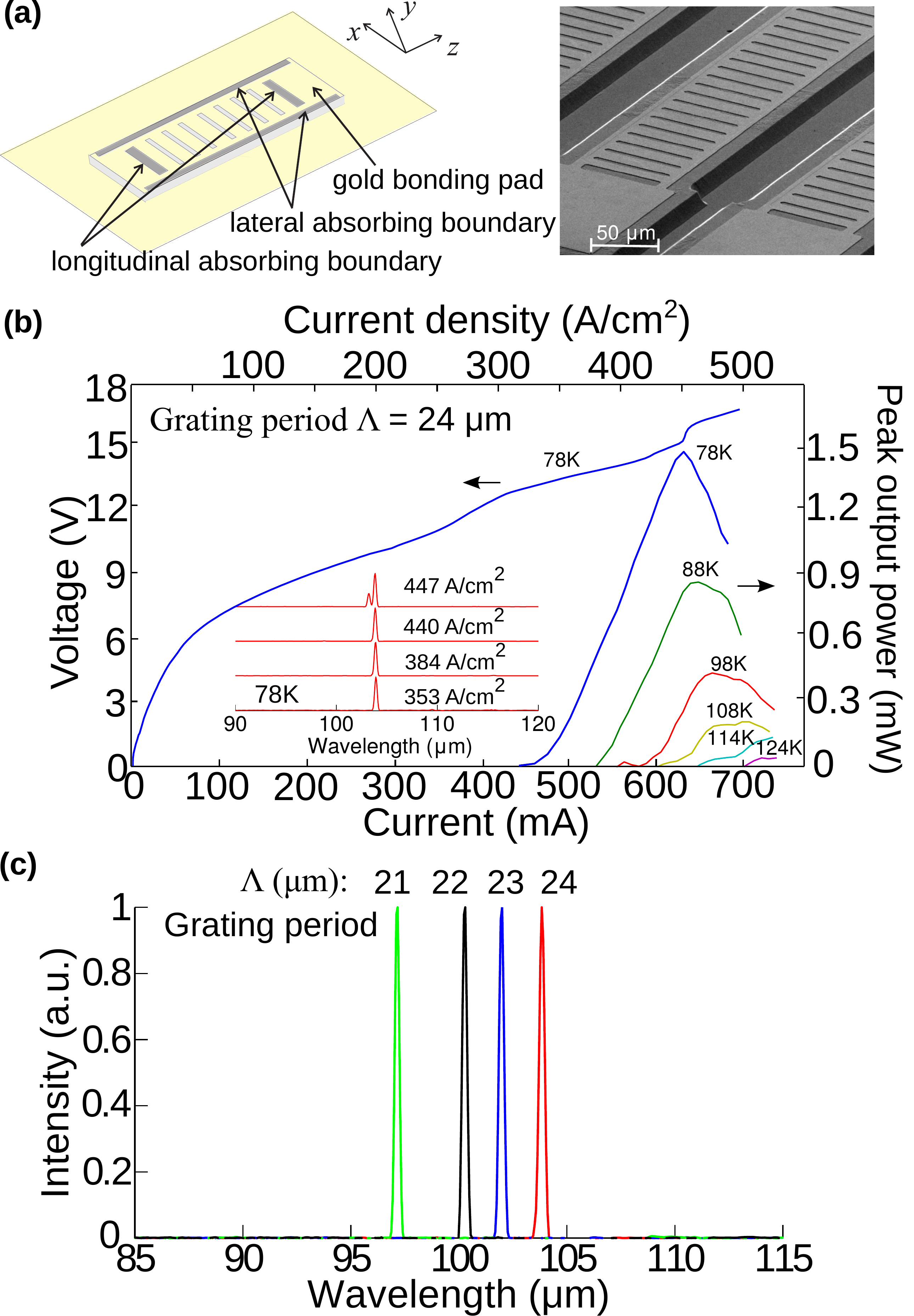}
\vspace{-0.10in}
\caption{
{\bf Lasing characteristics of terahertz QCLs with antenna-feedback.}
(a)~The schematic on left shows the QCL's metallic-cavity with antenna-feedback grating implemented
in top metal cladding. The
active-medium is $10~\um$ thick and based on a $3~\thz$ GaAs/\AGAt\ QCL design (details in supplementary material, section~S1).
A scanning-electron microscope image of the fabricated QCLs is shown on the right.
(b)~Experimental light-current-voltage characteristics of a representative QCL with antenna-feedback of 
dimensions $1.4~\rmmm\times 100~\um$ at different heat-sink temperatures.
The QCL is biased with low duty-cycle current pulses of $200~$ns duration and $100~$kHz
repetition rate. Inset shows lasing spectra for different bias where the spectral linewidth is limited
by instrument's resolution. The emitted optical power is measured without any cone collecting optic inside the cryostat.
(c)~Measured spectra for four different antenna-feedback QCLs with varying grating periods $\Lambda$, but similar overall
cavity dimensions. The QCLs are biased at a current-density $\sim 440~\Appcm$ at $78~\rmK$.
\vspace{-0.1in}
}
\label{fig3}
\end{figure}

A third-order DFB technique was recently shown to achieve emission in a narrow beam for terahertz QCLs with
\FPerot\ cavities~\cite{amanti:dfb}. It can achieve high directionality
for the radiated beam in both directions perpendicular to propagation, so long as the effective propagation index
of the SPP wave inside the active medium could be made $\sim 3.0$ by complex deep dry etching in the slits ~\cite{amanti:dfb}
or lateral corrugated geometry ~\cite{amanti:dfbcw}. The so-called phase matching condition
is possible for GaAs-based QCLs by cavity engineering~\cite{kao:dfbpm} since the $n_\rmGaAs\sim 3.6$ is
close to $3.0$. The antenna-feedback technique in this work offers similar outcome as a perfectly matched
third-order DFB with improved directionality as well as a novel outcoupling mechanism of the radiated beam
from terahertz QCLs. It is to be noted that the antenna-feedback scheme is automatically phase-matched and hence it
could be utilized for any type of spaser without any restrictions on the required index in the active medium. 

\section{Simulation results}
\label{Simulation}

Figure~\ref{fig2} shows comparison of the eigenmode spectrum of a terahertz QCL cavity with conventional DFB, taking third-order
DFB as an example versus antenna-feedback gratings computed using a finite-element solver~\cite{comsol:ref}.
The occurrence
of band-gaps in the spectra is indicative of the DFB effect. In both cases, the lower-frequency band-edge
mode is the lowest-loss mode by way of DFB action, since the DFB modes result in a standing-wave being
established along the length of the cavity with an envelope shape that vanishes close to the longitudinal
boundaries (end-facets). For antenna-feedback grating, a standing-wave for the single-sided SPP wave
is additionally established in air, as can be seen from the field-plot of the band-edge mode in
Fig.~\ref{fig2}(b). In contrast, third-order DFB leads to negligible amplitude of the single-sided SPP
wave in air, as mentioned previously and illustrated schematically in Fig.~\ref{fig1}(c). For the dominant TM polarized (Ey) electric field of antenna-feedback, the hybrid SPPs mode bound to the top metal layer consists of both quasi-cylindrical waves and SPPs, which are evanescent-field with a free-space propagation constant. Particularly at long wavelengths, such as mid infrared and THz region, SPPs and quasi-cylindrical waves complexly mix with each other~\cite{lalanne:creepingwave2}, contributing to the large spatial extent of SPP mode in surrounding medium as shown in Fig.~\ref{fig2}(b), which does not exist for any conventional DFB, such as first-order, second-order and third-order DFB. In addition, hybrid SPPs on top of metallic grating and standing-wave inside laser cavity show clearly different periodicity, with the ratio of free space wavelength over guided wavelength, which further confirms that the excitation of the coherent SPP wave on both side of top metallic surface contribute to feedback and coupling mechanism with antenna-feedback scheme. 
The absorbing boundaries at the longitudinal ends of the cavity~\cite{olivier:plasmon}
increase the relative loss of the modes that are further away from the band-edge mode, which helps
in mode discrimination and will lead to excitation of the desired band-edge mode for single-mode operation
of the spaser. The active region and metal layers are modeled as lossless since the exact
loss contribution due to each is not clear in literature for terahertz QCLs at cryogenic temperatures.
If lossy metal is used, the relative loss of various resonant modes for the DFB cavities are not impacted and
neither are the mode-shapes and the corresponding resonant frequencies. 
For the band-edge modes in Fig.~\ref{fig2}(a) and Fig.~\ref{fig2}(b), a loss of $\sim 5~\icm$
was estimated as a contribution from the absorbing boundaries. Consequently, the radiative (outcoupling) loss of
of the third-order DFB is $\sim 1.7~\icm$ as compared to $\sim 5.6~\icm$ for the antenna-feedback.
The radiative loss of the third-order DFB is smaller since the band-edge mode
has zeros of the radiative field ($E_z$) being located at each aperture, since the
grating period $\Lambda$ is integer multiple of half-wavelengths in the GaAs/AlGaAs
active medium ($\Lambda=3\lambda_\rmGaAs/2$, where $\lambda_\rmGaAs\equiv \lambda/n_\rmGaAs$). Such a
low-outcoupling efficiency is also existent in surface-emitting terahertz QCLs with second-order
DFB~\cite{kumar:surfemit}. For the cavity with antenna-feedback, the radiative loss is higher because the grating
period is not an integer multiple of half-wavelengths inside the active medium ($\Lambda\sim 0.78\lambda_\rmGaAs$)
that leads to large amplitudes of the radiative-field ($E_z$) in alternating apertures as shown
in the figure. As a consequence, the output power from terahertz QCLs with antenna-feedback should be
greater than that with conventional DFB gratings, which is an additional advantage of the antenna-feedback scheme
for terahertz QCLs. This was also verified experimentally from the measured output power.
The recently developed second-order DFB QCLs with graded periodicity~\cite{xu:aperiodic} 
achieve high-power emission for the same reason, \ie\ non-zero radiative field under
the metallic apertures. 

\section{Experimental demonstration of antenna-feedback for TH\lowercase{z} QCLs}
\label{Experimental}

Figure~\ref{fig3} shows experimental results from terahertz QCLs implemented with antenna-feedback
gratings. Details about fabrication and measurement methods are presented in supplementary material (section~S1).
Fig.~\ref{fig3}(b) shows representative \LI\ curves
versus heat-sink temperature for a QCL with $\Lambda=24~\um$. The QCL operated up to a temperature
of $124~\rmK$. In comparison, multi-mode \FPerot\ cavity QCLs on the same chip that did not include
longitudinal or lateral absorbing boundaries operated up to $\sim 140~\rmK$. Light-current characteristics and
spectra at different bias with \FPerot\ cavity are shown in supplementary material (section~S3). The
temperature degradation due to absorbing boundaries is relatively small and similar to previous reports
of DFB terahertz QCLs~\cite{kumar:surfemit}. The inset shows measured spectra at different bias at $78~\rmK$.
Most QCLs tested with different grating periods showed robust single-mode operation except close to peak-bias
when a second-mode was excited for some devices at a shorter-wavelength, which suggests it is likely due to
a higher-order lateral mode being excited due to spatial-hole burning in the cavity. Peak-power
output of $\sim 1.5~\rmmW$ was detected from the antenna-feedback QCL measured directly at the detector
without using any collecting optics.
For comparison, a terahertz QCL with third-order DFB (without phase matching) and similar dimensions was also
fabricated on the same chip, which operated up to a similar temperature of $\sim 124~\rmK$ and emitted peak-power
output of $\sim 0.45~\rmmW$ (see supplementary material, section~S3). The antenna-feedback
gratings lead to greater radiative outcoupling compared to conventional DFB schemes
for terahertz QCLs, as discussed in the previous section.

Fig.~\ref{fig3}(c) shows spectra measured from four different terahertz QCLs
with antenna-feedback gratings of different grating periods $\Lambda$. The single-mode 
spectra scales linearly with $\Lambda$, which is the clearest proof that the feedback
mechanism works as expected and the lower band-edge mode is selectively excited in each case.
Using $\lambda=\Lambda(n_\rma+1)$ from equation~\eqref{eq2}, 
the effective propagation index of the SPP mode in the active medium $n_\rma$ is calculated
as $3.59$, $3.53$, $3.46$, and $3.33$ for QCLs with $\Lambda$ of $21~\um$, $22~\um$, $23~\um$,
and $24~\um$ respectively.
The effective mode-index $n_\rma$ decreases because a larger $\Lambda$ introduces larger
sized apertures in the metal film since the grating duty-cycle was kept same for all devices.
Consequently, a greater amount of field couples to the single-sided SPP mode in air for
increasing $\Lambda$, thereby reducing the modal confinement in the active medium that reduces
the propagation index of the guided SPP mode further.

\begin{figure}[htbp]
\centering
\includegraphics[width=3.3in]{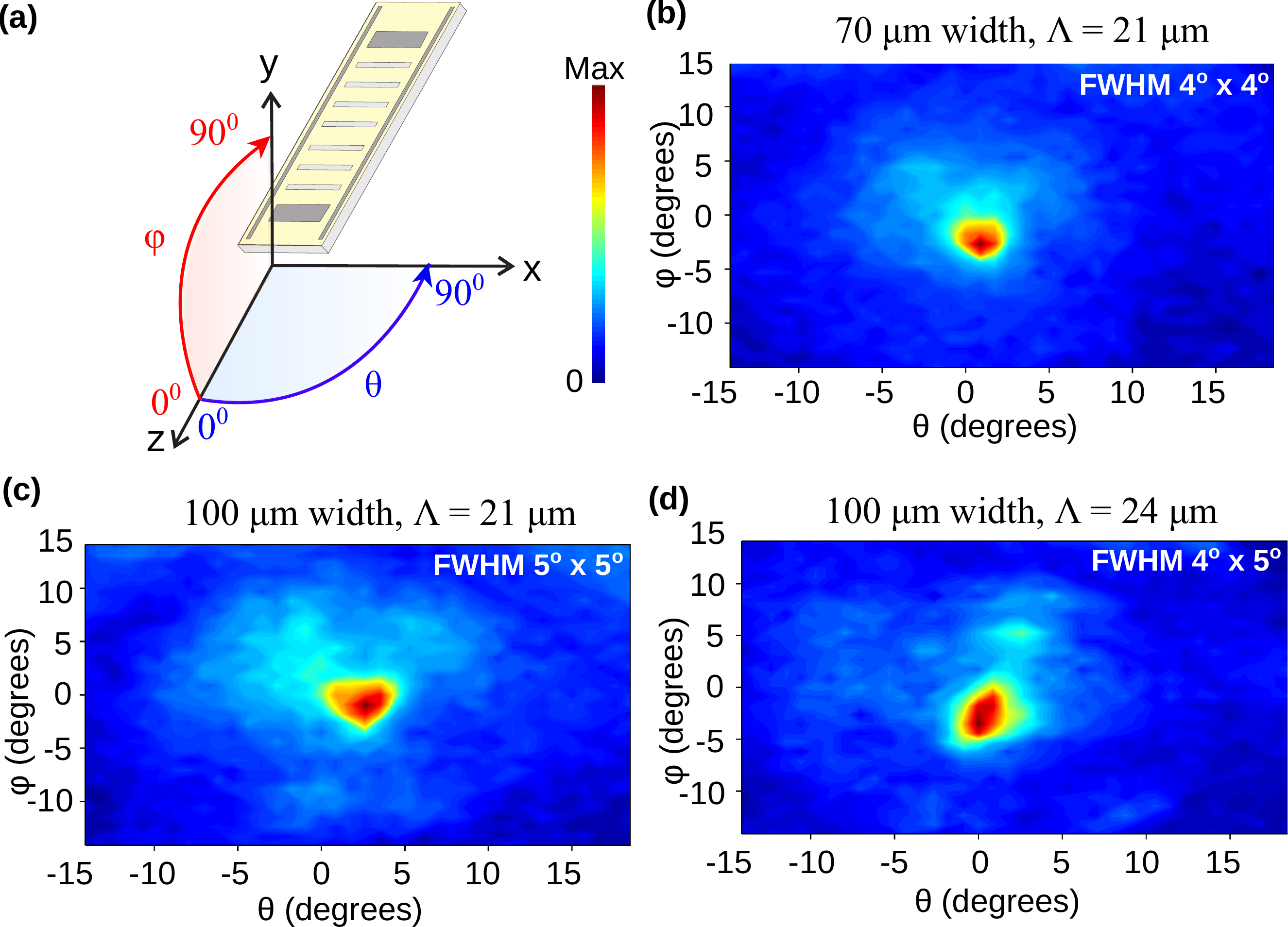}
\vspace{-0.10in}
\caption{
{\bf Far-field radiation-patterns of terahertz QCLs with antenna-feedback.}
(a)~Schematic showing orientation of QCLs and definition of angles.
The QCLs were operated at $78~\rmK$ in pulsed mode and biased at $\sim 440~\Appcm$ while lasing in single-mode.
The plots are for QCLs with
$\sim 1.4~$mm long cavities and (b)~$70~\um$ width and $\Lambda=21~\um$ grating emitting at $\sim 3.1~\thz$,
(c)~$100~\um$ width and $\Lambda=21~\um$ grating emitting at $\sim 3.1~\thz$, and (d)~$100~\um$ width and $\Lambda=24~\um$ grating
emitting at $\sim 2.9~\thz$ respectively.
\vspace{-0.1in}
}
\label{fig4}
\end{figure}

Experimental far-field beam patterns for antenna-feedback QCLs with varying designed parameters
are shown in Fig.~\ref{fig4}. Single-lobed beams in both lateral ($x$) and vertical ($y$) directions
were measured for all QCLs. As shown in Fig.~\ref{fig4}(b), the full-width half-maximum (FWHM) for the QCL
with $70~\um$ width, $\Lambda=21~\um$ is $\sim 4^{\circ} \times 4^{\circ}$, which is the narrowest reported
beam-profile from any terahertz QCL to-date.
In contrast, previous schemes for emission in a narrow-beam have resulted in divergence angles of
$6^\circ\times 11^\circ$ using very long ($>5~\rmmm$) cavities and a phased-matched third-order DFB scheme~\cite{kao:dfbpm} and
$7^\circ\times 10^\circ$ using broad-area devices with 2D photonic-crystals~\cite{halioua:phocrystal} for
single-mode terahertz QCLs, and $4^\circ\times10^\circ$~\cite{liang:collimated} and $12^\circ\times 16^\circ$~\cite{yu:spoof}
for multi-mode QCLs using metamaterial collimators.
Figs.~\ref{fig4}(c) and \ref{fig4}(d) show representative beam patterns from QCLs with wider cavities of $100~\um$ width, for the smallest and
largest $\Lambda$ in the range of fabricated devices respectively.
The beam divergence is relatively independent of $\Lambda$ as expected.
More importantly, the measurements show that the wider cavities result in a slightly broader
beam. Such a result is counter-intuitive because typically a laser emits in a narrower beam as 
its cavity's dimensions are increased due to an increase in the size of emitting aperture.
Such a behavior is unique for a spaser with antenna-feedback, and is discussed along with
full-wave 3D FEM simulation of the beam pattern in supplementary material (section~S2).
It can be argued that the size of the beam could be further narrowed by 
utilizing narrower cavities for terahertz QCLs, which will be extremely beneficial to
develop cw sources of narrow-beam coherent terahertz radiation.

\section{Conclusion}
\label{Conclusion}

In this article, we have presented a novel antenna-feedback scheme to achieve single-mode operation
and a highly directional far-field radiation pattern from plasmonic lasers with
subwavelength apertures and \FPerot\ type cavities.
It is conceptually different from any other previously utilized 
DFB schemes for solid-state lasers, and is based 
on phase-locking of a single-sided surface-plasmon-polariton (SPP) mode
on (one of) the metal film(s) in the spaser's cavity with
the guided SPP mode inside the spaser's active medium. 
The phase-locking is established due to strong Bragg diffraction of the 
SPP modes by periodically perforating the metal film in the form of a grating
of holes or slits. 
The uniqueness of the method lies in the specific value of the grating's period,
which leads to the spaser's cavity radiating like an end-fire phased-array antenna for
the excited DFB mode. Additionally, coherent single-sided
SPPs are also generated on the metal film that have a large spatial extent in the surrounding medium of the 
laser's cavity, which could have important implications for applications in integrated plasmonics.
Coherent SPPs with large spatial extent could make it easy to couple 
SPP waves from the plasmonic lasers to other photonic components, and could also potentially
be utilized for plasmonic sensing. Experimentally, the scheme is implemented in
terahertz QCLs with subwavelength metallic cavities. A beam-divergence angle as small as
$4^\circ\times 4^\circ$ is achieved for single-mode QCLs, which is narrower
than that achieved with any other previously reported schemes for terahertz QCLs with
periodic photonic structures.
Compared with the third-order DFB method, the new antenna-feedback scheme is easier to implement
for fabrication by standard lithography techniques without any other complex fabrication technique to precisely match a well defined effective mode index, and achieves a superior radiative outcoupling
owing to the fact that the grating period is not an integer multiple
of half-wavelengths of the standing SPP-wave inside the active medium. 
Terahertz QCLs with antenna-feedback could lead to
development of new modalities for terahertz spectroscopic sensing and wavelength
tunability due to access of a coherent terahertz SPP wave on top of the QCL's cavity,
possibilities of sensing and imaging at standoff distances of few tens of meters, and 
development of integrated terahertz laser arrays with a broad spectral coverage
for applications in terahertz absorption spectroscopy.


\subsection*{Supplementary Materials}
\vspace{-0.10in}

See supplementary materials section for additional supporting content.




\renewcommand\bibsection{\section*{\sc \large References}}


\begin{thebibliography}{10}
\newcommand{\enquote}[1]{``#1''}

\bibitem{bergman:spaser}
D.~J. Bergman and M.~I. Stockman, \enquote{Surface plasmon amplification by
  stimulated emission of radiation: Quantum generation of coherent surface
  plasmons in nanosystems,} Phys. Rev. Lett. \textbf{90}, 027402 (2003).

\bibitem{hill:smalllasers}
M.~T. Hill and M.~C. Gather, \enquote{Advances in small lasers,} Nat. Photon.
  \textbf{8}, 908 (2014).

\bibitem{oulton:splasers}
R.~F. Oulton, \enquote{Surface plasmon lasers: sources of nanoscopic light,}
  Mater. Today \textbf{15}, 26 (2012).

\bibitem{berini:spplasers}
P.~Berini and I.~D. Leon, \enquote{Surface plasmon-polariton amplifiers and
  lasers,} Nat. Photon. \textbf{6}, 16 (2012).

\bibitem{noginov:spaser}
M.~A. Noginov, G.~Zhu, A.~M. Belgrave, R.~Bakker, V.~M. Shalaev, E.~E.
  Narimanov, S.~Stout, E.~Herz, T.~Suteewong, and U.~Wiesner,
  \enquote{Demonstration of a spaser-based nanolaser,} Nature \textbf{460},
  1110 (2009).

\bibitem{kohler:laser}
R.~K\"{o}hler, A.~Tredicucci, F.~Beltram, H.~E. Beere, E.~H. Linfield, A.~G.
  Davies, D.~A. Ritchie, R.~C. Iotti, and F.~Rossi, \enquote{Terahertz
  semiconductor-heterostructure laser,} Nature \textbf{417}, 156--159 (2002).

\bibitem{williams:review}
B.~S. Williams, \enquote{Terahertz quantum-cascade lasers,} Nat. Photon.
  \textbf{1}, 517--525 (2007).

\bibitem{hill:mimlaser}
M.~T. Hill, M.~Marell, E.~S.~P. Leong, B.~Smalbrugge, Y.~Zhu, M.~Sun, P.~J. van
  Veldhoven, E.~J. Geluk, F.~Karouta, Y.-S. Oei, R.~N\"{o}tzel, C.-Z. Ning, and
  M.~K. Smit, \enquote{Lasing in metal-insulator-metal sub-wavelength plasmonic
  waveguides,} Opt. Express \textbf{17}, 11107 (2009).

\bibitem{oulton:plasmonlaser}
R.~F. Oulton, V.~J. Sorger, T.~Zentgraf, R.-M. Ma, C.~Gladden, L.~Dai,
  G.~Bartal, and X.~Zhang, \enquote{Plasmon lasers at deep subwavelength
  scale,} Nature \textbf{461}, 629 (2009).

\bibitem{lu:nanolaser}
Y.-J. Lu, J.~Kim, H.-Y. Chen, C.~Wu, N.~Dabidian, C.~E. Sanders, C.-Y. Wang,
  M.-Y. Lu, B.-H. Li, X.~Qiu, W.-H. Chang, L.-J. Chen, G.~Shvets, C.-K. Shih,
  and S.~Gwo, \enquote{Plasmonic nanolaser using epitaxially grown silver
  film,} Science \textbf{337}, 450 (2012).

\bibitem{williams:metal}
B.~S. Williams, S.~Kumar, H.~Callebaut, Q.~Hu, and J.~L. Reno,
  \enquote{Terahertz quantum-cascade laser at $\lambda \approx 100~\mu$m using
  metal waveguide for mode confinement,} Appl. Phys. Lett. \textbf{83},
  2124--2126 (2003).

\bibitem{adam:beam}
A.~J.~L. Adam, I.~Ka{\v s}alynas, J.~N. Hovenier, T.~O. Klaassen, J.~R. Gao,
  E.~E. Orlova, B.~S. Williams, S.~Kumar, Q.~Hu, and J.~L. Reno, \enquote{Beam
  patterns of terahertz quantum cascade lasers with subwavelength cavity
  dimensions,} Appl. Phys. Lett. \textbf{88}, 151105 (2006).

\bibitem{orlova:wirelaser}
E.~E. Orlova, J.~N. Hovenier, T.~O. Klaassen, I.~Ka{\v s}alynas, A.~J.~L. Adam,
  J.~R. Gao, T.~M. Klapwijk, B.~S. Williams, S.~Kumar, Q.~Hu, and J.~L. Reno,
  \enquote{Antenna model for wire lasers,} Phys. Rev. Lett. \textbf{96}, 173904
  (2006).

\bibitem{zhou:nanocavityarrays}
W.~Zhou, M.~Dridi, J.~Y. Suh, C.~H. Kim, D.~T. Co, M.~R. Wasielewski, G.~C.
  Schatz, and T.~W. Odom, \enquote{Lasing action in strongly coupled plasmonic
  nanocavity arrays,} Nat. Nanotech. \textbf{8}, 506 (2013).

\bibitem{beijnum:holearrays}
F.~van Beijnum, P.~J. van Veldhoven, E.~J. Geluk, M.~J.~A. de~Dood, G.~W. ’t
  Hooft, and M.~P. van Exter, \enquote{Surface plasmon lasing observed in metal
  hole arrays,} Phys. Rev. Lett. \textbf{110}, 206802 (2013).

\bibitem{meng:spaserarray}
X.~Meng, J.~Liu, A.~V. Kildishev, and V.~M. Shalaev, \enquote{Highly
  directional spaser array for the red wavelength region,} Laser \& Photon.
  Rev. \textbf{8}, 896 (2014).

\bibitem{yang:real}
A.~Yang, T.~B. Hoang, M.~Dridi, C.~Deeb, M.~H. Mikkelsen, G.~C. Schatz, and
  T.~W. Odom, \enquote{Real-time tunable lasing from plasmonic nanocavity
  arrays,} Nature Comm. \textbf{6}, 6939 (2015).

\bibitem{schokker:lasing}
A.~H. Schokker and A.~F. Koenderink, \enquote{Lasing at the band edges of
  plasmonic lattices,} Phys. Rev. B \textbf{90}, 155452 (2014).

\bibitem{dorofeenko:steady}
A.~V. Dorofeenko, A.~A. Zyablovsky, A.~P. Vinogradov, E.~S. Andrianov, A.~A.
  Pukhov, and A.~A. Lisyansky, \enquote{Steady state superradiance of a
  {2D-spaser} array,} Opt. Express \textbf{21}, 14539--14547 (2013).

\bibitem{mahler:review11}
L.~Mahler and A.~Tredicucci, \enquote{Photonic engineering of surface-emitting
  terahertz quantum cascade lasers,} Laser \& Photon. Rev. \textbf{5}, 647--658
  (2011).

\bibitem{sirtori:review13}
C.~Sirtori, S.~Barbieri, and R.~Collombelli, \enquote{Wave engineering with
  {THz} quantum cascade lasers,} Nat. Photon. \textbf{7}, 691 (2013).

\bibitem{kao:phase}
T.-Y. Kao, Q.~Hu, and J.~L. Reno, \enquote{Phase-locked arrays of
  surface-emitting terahertz quantum-cascade lasers,} Appl. Phys. Lett.
  \textbf{96}, 101106 (2010).

\bibitem{halioua:phaselocked}
Y.~Halioua, G.~Xu, S.~Moumdji, L.~Li, J.~Zhu, E.~H. Linfield, A.~G. Davies,
  H.~E. Beere, D.~A. Ritchie, and R.~Colombelli, \enquote{Phase-locked arrays
  of surface-emitting graded-photonic-heterostructure terahertz semiconductor
  lasers,} Opt. Express \textbf{23}, 6915 (2015).

\bibitem{xu:metasurface}
L.~Xu, C.~A. Curwen, P.~W. Hon, Q.-S. Chen, T.~Itoh, and B.~S. Williams,
  \enquote{Metasurface external cavity laser,} Appl. Phys. Lett. \textbf{107},
  221105 (2015).

\bibitem{amanti:dfb}
M.~I. Amanti, M.~Fischer, G.~Scalari, M.~Beck, and J.~Faist,
  \enquote{Low-divergence single-mode terahertz quantum cascade laser,} Nat.
  Photon. \textbf{3}, 586--590 (2009).

\bibitem{kao:dfbpm}
T.~Y. Kao, Q.~Hu, and J.~L. Reno, \enquote{Perfectly phase-matched third-order
  {DFB THz} quantum-cascade lasers,} Opt. Lett. \textbf{37}, 2070 (2012).

\bibitem{williams:corrdfb}
B.~S. Williams, S.~Kumar, Q.~Hu, and J.~L. Reno, \enquote{Distributed-feedback
  terahertz quantum-cascade lasers with laterally corrugated metal waveguides,}
  Opt. Lett. \textbf{30}, 2909 (2005).

\bibitem{fan:surfemit}
J.~A. Fan, M.~A. Belkin, F.~Capasso, S.~Khanna, M.~Lachab, A.~G. Davies, and
  E.~H. Linfield, \enquote{Surface emitting terahertz quantum cascade laser
  with a double-metal waveguide,} Opt. Express \textbf{14}, 11672 (2007).

\bibitem{kumar:surfemit}
S.~Kumar, B.~S. Williams, Q.~Qin, A.~W.~M. Lee, Q.~Hu, and J.~L. Reno,
  \enquote{Surface-emitting distributed feedback terahertz quantum-cascade
  lasers in metal-metal waveguides,} Opt. Express \textbf{15}, 113 (2007).

\bibitem{chassagneux:nature}
Y.~Chassagneux, R.~Colombelli, W.~Maineult, S.~Barbieri, H.~E. Beere, D.~A.
  Ritchie, S.~P. Khanna, E.~H. Linfield, and A.~G. Davies,
  \enquote{Electrically pumped photonic-crystal terahertz lasers controlled by
  boundary conditions,} Nature \textbf{457}, 174 (2009).

\bibitem{halioua:phocrystal}
Y.~Halioua, G.~Xu, S.~Moumdji, L.~H. Li, A.~G. Davies, E.~H. Linfield, and
  R.~Colombelli, \enquote{{THz} quantum cascade lasers operating on the
  radiative modes of a {2D} photonic crystal,} Opt. Lett. \textbf{39}, 3962
  (2014).

\bibitem{lalanne:creepingwave2}
P.~Lalanne, J.~Hugonin, H.~Liu, and B.~Wang, \enquote{A microscopic view of the
  electromagnetic properties of sub-$\lambda$ metallic surfaces,} Surf. Sci.
  Rep. \textbf{64}, 453 (2009).

\bibitem{babuty:mirspsource}
A.~Babuty, A.~Bousseksou, J.-P. Tetienne, I.~M. Doyen, C.~Sirtori, G.~Beaudoin,
  I.~Sagnes, Y.~D. Wilde, and R.~Colombelli, \enquote{Semiconductor surface
  plasmon sources,} Phys. Rev. Lett. \textbf{104}, 226806 (2010).

\bibitem{amanti:dfbcw}
M.~I. Amanti, G.~Scalari, F.~Castellano, M.~Beck, and J.~Faist, \enquote{Low
  divergence terahertz photonic-wire laser,} Opt. Express \textbf{18}, 6390
  (2010).

\bibitem{comsol:ref}
{COMSOL 4.4}, a finite-element partial differential equation solver from COMSOL
  Inc.

\bibitem{olivier:plasmon}
O.~Demichel, L.~Mahler, T.~Losco, C.~Mauro, R.~Green, A.~Tredicucci, J.~Xu,
  F.~Beltram, H.~E. Beere, D.~A. Ritchie, and V.~Tamo{\v s}inuas,
  \enquote{Surface plasmon photonic structures in terahertz quantum cascade
  lasers,} Opt. Express \textbf{14}, 5335 (2006).

\bibitem{xu:aperiodic}
G.~Xu, R.~Colombelli, S.~P. Khanna, A.~Belarouci, X.~Letartre, L.~Li, E.~H.
  Linfield, A.~G. Davies, H.~E. Beere, and D.~A. Ritchie, \enquote{Efficient
  power extraction in surface-emitting semiconductor lasers using graded
  photonic heterostructures,} Nature Comm. \textbf{3}, 952 (2012).

\bibitem{liang:collimated}
G.~Liang, E.~Dupont, S.~Fathololoumi, Z.~R. Wasilewski, D.~Ban, H.~K. Liang,
  Y.~Zhang, S.~F. Yu, L.~H. Li, A.~G. Davies, E.~H. Linfield, H.~C. Liu, and
  Q.~J. Wang, \enquote{Planar integrated metasurfaces for highly-collimated
  terahertz quantum cascade lasers,} Sci. Rep. \textbf{4}, 7083 (2014).

\bibitem{yu:spoof}
N.~Yu, Q.~J. Wang, M.~A. Kats, J.~A. Fan, S.~P. Khanna, L.~Li, A.~G. Davies,
  E.~H. Linfield, and F.~Capasso, \enquote{Designer spoof surface plasmon
  structures collimate terahertz laser beams,} Nat. Materials \textbf{9}, 730
  (2010).

\bibitem{khanal:rpreview}
S.~Khanal, L.~Zhao, J.~L. Reno, and S.~Kumar, \enquote{Temperature performance
  of terahertz quantum-cascade lasers with resonant phonon active-regions,} J.
  Opt \textbf{16}, 094001 (2014).

\bibitem{balanis:antenna}
C.~A. Balanis, \emph{Antenna Theory: Analysis and Design} (Wiley-Interscience,
  2005), 3rd ed.

\bibitem{boyle:high}
C.~Boyle, C.~Sigler, J.~Kirch, D.~Lindberg~III, T.~Earles, D.~Botez, and
  L.~Mawst, \enquote{High-power, surface-emitting quantum cascade laser
  operating in a symmetric grating mode,} Appl. Phys. Lett. \textbf{108},
  121107 (2016).


\end{thebibliography}

\vspace{0.05in}

\footnotesize
\section*{Funding Information}
National Science Foundation (NSF) (ECCS 1128562, ECCS 1351142, CMMI 1437168).

\section*{Acknowledgments}
This work is performed, in part, at the Center for Integrated
Nanotechnologies, a U.S. Department of Energy (DOE), Office of Basic Energy Sciences user facility. Sandia National Laboratories
is a multiprogram laboratory managed and operated by Sandia Corporation, a wholly owned subsidiary of Lockheed Martin Corporation,
for the U.S. DOE's National Nuclear Security Administration under contract DE-AC04-94AL85000.


\section*{Competing financial interests}

A United States patent application (pending) for the described technology has been filed through Lehigh University (application number 14/984,652,
filed on Dec. 30, 2015).

\newpage
\clearpage

\setcounter{figure}{0}
\setcounter{section}{0}
\setcounter{subsection}{0}

\renewcommand{\thefigure}{S\arabic{figure}}
\renewcommand{\thesection}{S\arabic{section}}
\renewcommand{\thesubsection}{S\arabic{subsection}}

\makeatletter
\def\p@subsection{}
\makeatother

\setcounter{page}{1}

\Large
{\centering Terahertz plasmonic laser radiating in an ultra-narrow beam\\}
\large
{\centering C.~Wu \etal\\}

\section*{\Large Supplementary Materials}
\normalsize

\section{Materials and Methods}
\label{materials}

\subsection*{Fabrication and measurement details}
\label{fabrication}

The active-medium of the QCLs is based on a three-well resonant-phonon design with GaAs/\AGAt\ superlattice
(design RTRP3W197, wafer number VB0464), which is described in Ref.~\onlinecite{khanal:rpreview}, and was grown by molecular-beam epitaxy.
The QCL superlattice is $10~\um$ thick with an average $n$-doping of $5.5\times 10^{15}~\iiicm$, and surrounded
by $0.05~\um$ and $0.1~\um$ thick highly-doped GaAs contact layers at $5\times 10^{18}~\iiicm$ on either side
of the superlattice. Fabrication of QCLs with parallel-plate metallic cavities followed a Cu-Cu thermocompression
wafer bonding technique as in Ref.~\onlinecite{kumar:surfemit} with standard optical contact lithography.
Following wafer-bonding and substrate removal, positive-resist lithography was used to selectively etch away the
$0.1~\um$ thick highly-doped GaAs layer from all locations where top-metal cladding would exist on individual cavities
except in $\sim 5~\um$ wide regions at the outer rectangular boundaries of the top-metal layer due to overlapping mask
layers. The removal of this layer beneath the top-metal does not impact electrical
transport significantly except adding a small voltage drop at the top contact during QCL operation.
Importantly, this lithography step allows radiative outcoupling from the apertures in the finally fabricated 
QCL cavities; and simultaneously serves to implement longitudinal and lateral absorbing boundaries
as illustrated in Fig.~3(a) by leaving the highly-doped GaAs layer exposed at both longitudinal and lateral edges of
the QCL cavity. The absorbing boundaries result in a highly lossy propagation of the SPP modes in those
regions~\cite{olivier:plasmon}. While the longitudinal boundaries help in DFB mode discrimination, the
lateral boundaries are useful to eliminate higher-order lateral guided modes by making them more
lossy in comparison to the fundamental mode~\cite{kumar:surfemit,chassagneux:nature,amanti:dfb}.

Ti/Au metal layers of thickness $25/200~$nm were used as the top metal cladding, using an image-reversal lithography mask
for implementing gratings in the metal layers. Another positive-resist lithography step was used to cover the grating-metal
with photoresist to be used as a mask for wet-etching of ridges in a H$_2$SO$_4$:H$_2$O$_2$:H$_2$O $1$:$8$:$80$ solution for
$\sim 22~$minutes. After etching the $10~\um$ thick superlattice active region when the bottom metal layer is exposed,
an over-etch of $\sim 1~$minute was followed up to reduce the slope on the sidewalls of the QCL's cavities to allow
for a more uniform current-density distribution through the height of the cavity (in $y$ direction). A Ti/Au contact was used
as the backside-metal contact for the finally fabricated QCL chips to assist in soldering.
Before deposition of backside-metal of the wafer, the substrate was mechanically polished down to
a thickness of $\sim 170~\um$ to improve heat-sinking.

\begin{figure*}[htbp]
\centering
\includegraphics[width=6.5in]{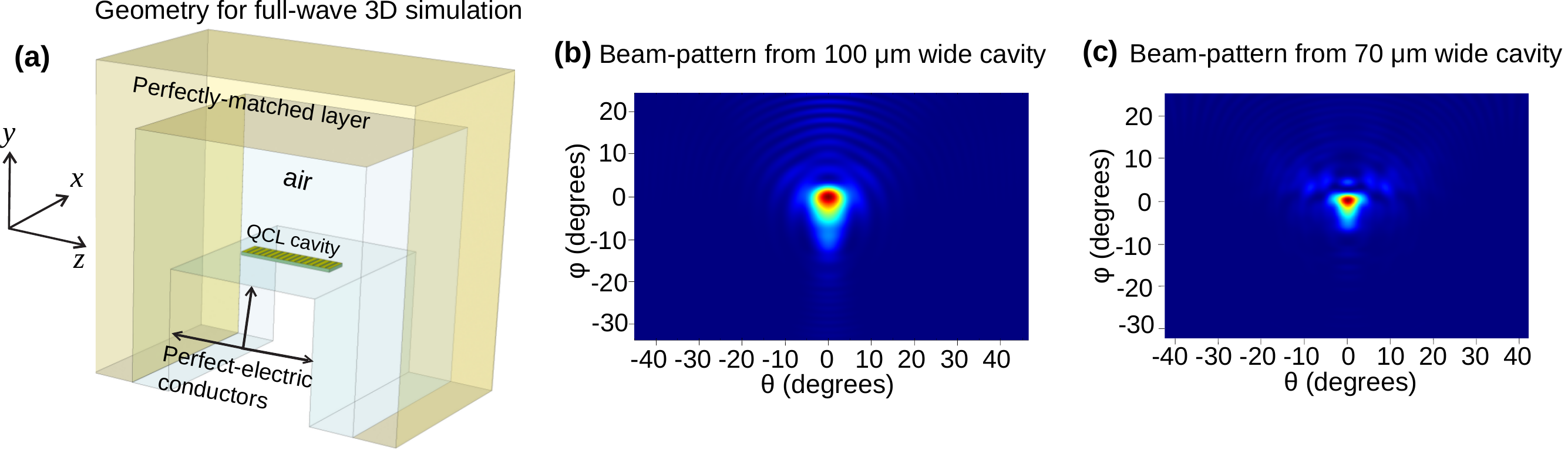}
\caption{
{\bf Full-wave 3D simulation to compute far-field radiation pattern of QCL cavities with antenna-feedback.}
(a)~The modeled geometry for full-wave 3D simulation with the FEM solver~\cite{comsol:ref}.
Parallel-plate metallic terahertz QCL cavities of $1.4~\rmmm$ length, $10~\um$ thickness, varying widths, and
a grating period of $\Lambda=21.7~\um$ were implemented, which excite antenna-feedback resonant-cavity modes at 
$\nu\sim 3.1~\thz$. (b) Simulated far-field radiation pattern of the cavity with $100~\um$ width. The FWHM is
$\sim 7^\circ \times 7^\circ$. (c) Far-field radiation pattern of the cavity with $70~\um$ width. The FWHM is
$\sim 6^\circ \times 5^\circ$.
}
\label{figS1}
\end{figure*}

For measurements, a cleaved chip consisting of QCLs with antenna-feedback gratings of different periods along with some \FPerot\
ridge QCLs and third-order DFB QCLs was In-soldered on a Cu block, the QCL to be tested was wire-bonded for electrical
biasing, and the Cu block was mounted on the cold-stage of a liquid Nitrogen vacuum cryostat for measurements.
Initial measurements of the fabricated QCL cavities did not yield lasing devices at $78~\rmK$. It was then realized that the
$\sim 13~\um$ wide lateral absorbing boundaries on each side were perhaps introducing high optical losses in the
cavities. The finally fabricated wafers were then etched again (post-fabrication) in a H$_2$SO$_4$:H$_2$O$_2$:H$_2$O $1$:$8$:$80$
solution for $\sim 15~$seconds, which is likely to etch away most of the exposed highly-doped GaAs 
layer. However, $\sim 5~\um$ wide highly-doped GaAs layer remains unetched underneath the top-metal cladding on all
four outer boundaries of the cladding, which serves to perform the role of an absorbing region. The post-etched devices
lased upon re-testing, and their robust single-mode operation suggests that both lateral and longitudinal absorbing boundaries
were able to suppress undesired resonant-modes in the cavity as desired.

The emitted optical power was measured with a deuterated triglycine sulfate pyroelectric detector (DTGS) pyroelectric detector
and calibrated with a terahertz thermopile powermeter (ScienTech model AC2500H) without any
optical component or cone collecting optic inside the cryostat to improve collection of radiated power in the case of
DFB QCLs. (A Winston cone was used to collect power, but only for the \FPerot-cavity QCL reported in Fig.~\ref{figS3}.)
The spectra were measured at $78~\rmK$ in linear-scan mode with a resolution of $0.2~\icm$ ($6$~GHz) using a Bruker Fourier-transform
infrared spectrometer (FTIR) under vacuum, with a room-temperature DTGS detector placed inside the FTIR.
Far-field radiation patterns were measured with a DTGS pyroelectric cell detector mounted on a computer-controlled two-axis
($x-y$ plane) moving stage in the end-fire ($z$) direction at a distance of $65~$mm from the QCLs' end-facets. The
detector's size leads to spatial averaging of the measured beam with a step-size of $1.7^\circ$, which has a
negligible impact on the measured beam-width.

\subsection*{Modeling}
\label{modeling}

Finite-element (FEM) simulations were performed using a commercially available software~\cite{comsol:ref}.
The eigenmodes of the DFB cavities are computed using an eigenfrequency solver. The loss in absorbing
boundaries of the cavity is implemented using a complex dielectric constant computed using Drude-model~\cite{kumar:surfemit}. 
The imaginary part of the computed eigenfrequencies is used to compute the propagation loss of the resonant-cavity
modes in the units of inverse of length. The active region is modeled as lossless and the metal is modeled to be perfect electrical conductors, in which
case, the computed loss is the sum of loss at absorbing boundaries as well as that due to radiation (outcoupling). The thickness of the metallic layer is $400~$nm in simulation. The relative losses of various resonant modes for the DFB cavities are not changed and neither are the mode-shapes and the corresponding resonant frequencies when using lossy metal, or when the thickness of the metallic film are altered (as long as it is kept significantly thinner than that of the active region). 
For the 2D simulations (cavities of infinite-width) the far-field boundaries
are modeled as absorbing layers as described in Ref.~\cite{kumar:surfemit}. For 3D simulations, the in-built
perfectly matched layers in the FEM solver were utilized to model the far-field boundaries.

\section{Full-wave 3D simulation for radiation-patterns}
\label{full}

The narrow-beam emission from plasmonic lasers with antenna-feedback is due to a combination
of two factors that are related to the radiative behavior of phased-array antennas.
In addition to the {\em array-factor} that leads to narrower beams for more
number of elements in a phased-array (\ie\ in this case, a longer length of the spaser's cavity),
the far-field radiation pattern is additionally narrowed owing to an {\em element-factor}, which
is defined as the far-field radiation pattern due to an individual emitter of the phased-array~\cite{balanis:antenna}.
The large spatial extent of the single-sided SPP wave in the surrounding medium results in a narrower
element-factor compared to an omnidirectional point-source emitter. 

As measured experimentally and shown in Fig.~4, the antenna-feedback terahertz QCL cavities with a
narrower width of $70~\um$ resulted in a beam with smaller divergence as compared to the $100~\um$ wide cavity.
This seemingly unique behavior from antenna-feedback QCLs serves to further
validate the concept of the specific feedback scheme, in which the cavity radiates like
a phased-array antenna. A narrower cavity causes greater lateral spread of the single-sided SPP mode
(in the $x$ dimension) in the surrounding medium, especially when the cavity's width is sub-wavelength.
A single-sided SPP mode with a broader cross-section in the $x-y$ plane will lead to a more directional
far-field radiation pattern owing to a narrower element-factor for the phased-array antenna structure.

To further validate the experimental results of Fig.~4, full-wave 3D FEM simulations were carried
out for terahertz QCL cavities with antenna-feedback. Figure~\ref{figS1} shows the computed far-field radiation patterns
for band-edge DFB mode for terahertz QCL cavities
implemented with antenna-feedback gratings in the top metal cladding. The in-built perfectly-matched layers (PMLs)
in the FEM solver~\cite{comsol:ref} serve as an effective absorbing boundary for computation of the radiation pattern.
The PMLs' parameters are coarsely adjusted to achieve minimum reflection at terahertz frequencies and transform the
propagating waves to exponentially decaying waves. The PMLs are placed far enough to avoid interaction with the
single-side SPP standing-wave in the near-field of the cavity. Considering the computer's memory limitations,
laterally absorbing boundaries are not modeled in the cavity, which does not impact the simulation's results
since the DFB mode could always be found, even though it is not the lowest loss mode in the eigenmode spectrum 
of the cavity due to other higher order lateral modes that can have even lower propagation loss. All metal
layers are modeled as perfectly conducting surfaces to limit the mesh size in the geometry.  
In order to set up a far-field calculation, a far-field domain node is implemented that
is a single closed surface surrounding all radiating apertures in the cavity. The distribution of far EM field
is based on the Fourier transform of the near-field as implemented in the FEM solver. Simulated 3D beam patterns
demonstrate single-lobed beams with narrow divergence for the band-edge mode with antenna-feedback as shown in 
the figure. The shape and FWHM of the computed radiation-pattern is in close agreement with the experimentally
measured beams that are shown in Fig.~4. The FWHM of simulated beam patterns is slightly larger than that
of the measured results. The difference are likely due to the relative simplification of implemented 3D model, which
does not account for the sloped sidewall profile of the ridges and also the fact the ground plane around the 
QCL's cavity is different for the measured chip as compared to that in the simulated geometry.

\begin{figure}[htbp]
\centering
\includegraphics[width=3.3in]{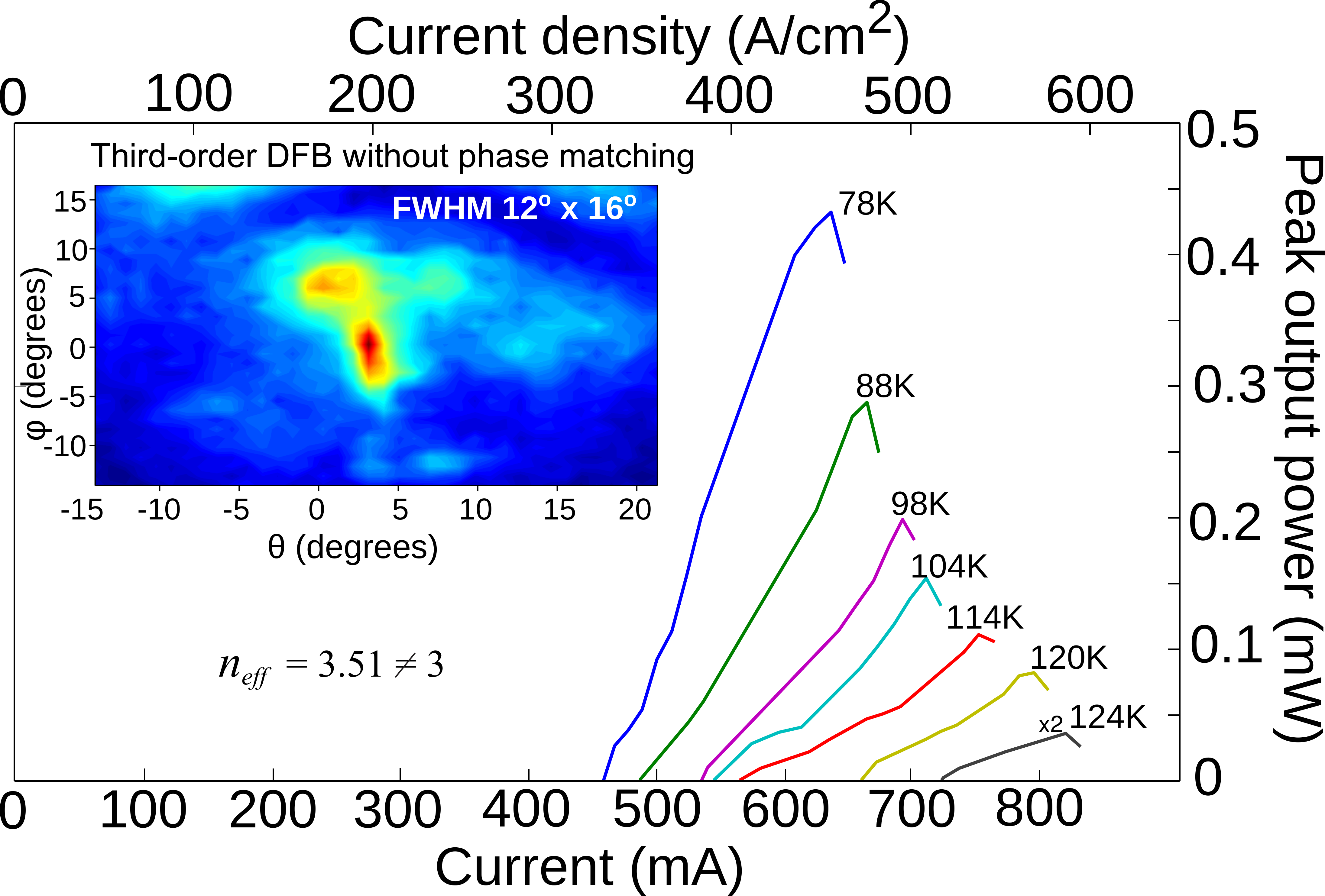}
\caption{
{\bf Lasing characteristics of a terahertz QCL with third-order DFB without phase matching.}
Light-current characteristics of a terahertz QCL with third-order DFB in pulsed operation. The QCL 
is fabricated on the same chip, and the cavity dimensions are similar to the QCLs with antenna-feedback 
for which data is presented in Fig.~3. The inset shows the measured far-field radiation-pattern
with the same angular definitions as in Fig.~4(a). The QCL emitted
predominantly in single-mode at $\lambda=103.0~\um$ ($\nu=2.91~\thz$) with grating period $44~\um$ .
}
\label{figS2}
\end{figure}

\begin{figure}[htbp]
\centering
\includegraphics[width=3.3in]{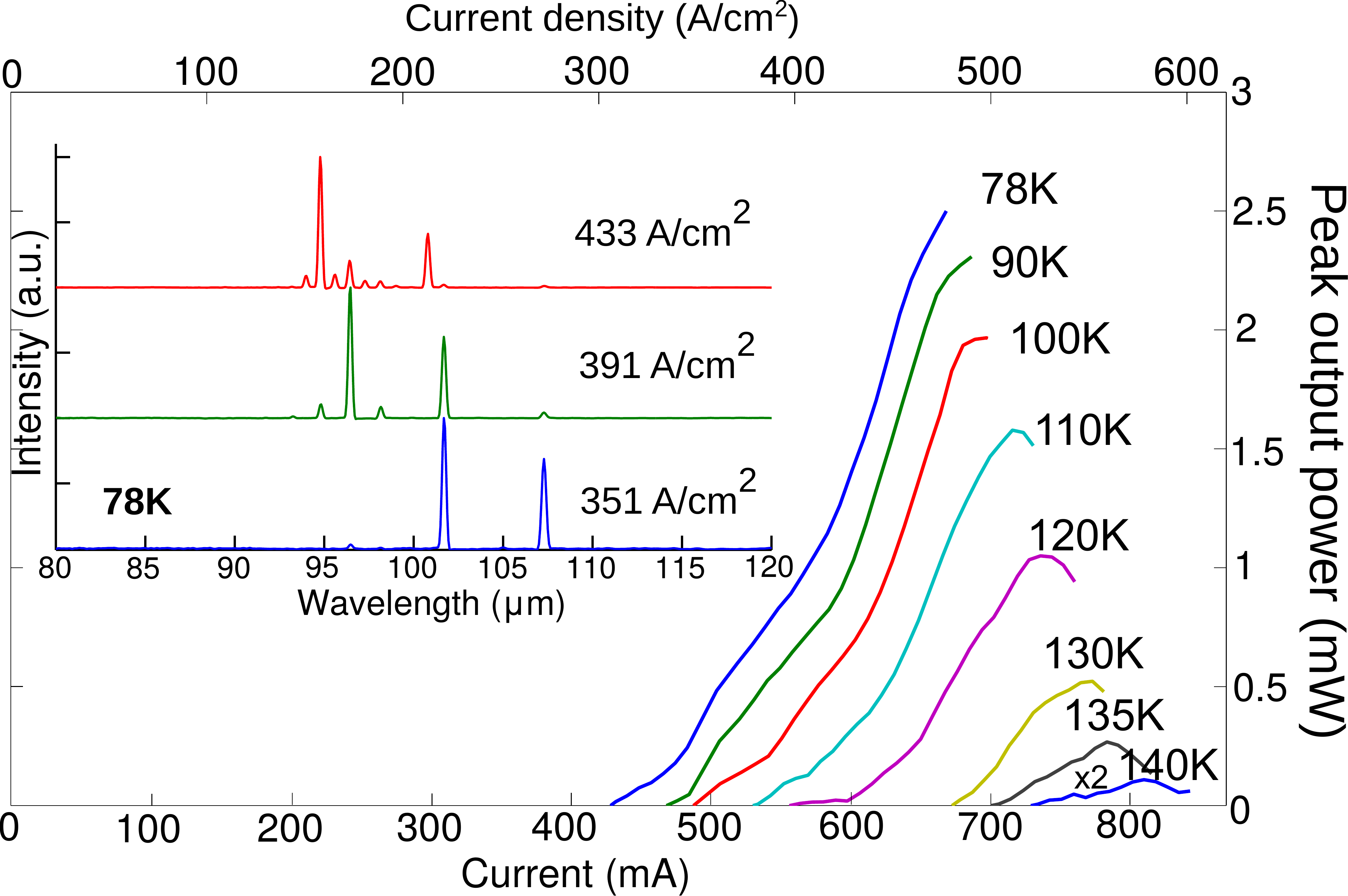}
\caption{
{\bf Lasing characteristics of a terahertz QCL with  \FPerot\ cavity.}
Light-current characteristics of a QCL with \FPerot\ cavity at different heat-sink temperatures in pulsed operation. Output power was collected with a Winston cone placed close to laser facet due to the highly divergent beam pattern of \FPerot\ type THz QCLs. The inset shows lasing spectra at different bias at $\sim 78~\rmK$. .
}
\label{figS3}
\end{figure}

\section{Experimental characteristics of third-order DFB QCL and a \FPerot-cavity QCL}
\label{comparison}

Figure~\ref{figS2} shows \LI\ characteristics of a representative terahertz QCL with third-order DFB without well defined phase-matching condition that was fabricated on the same chip by standard lithography to pattern grating on top metallic layer, and with similar dimensions as that of the antenna-feedback
QCLs whose data is reported in Fig.~3. Using the phase condition $\lambda=2 n_\rma \Lambda/p$
as determined from equation~1 for $p=3$ (third-order DFB), an effective-mode index
of $n_\rma=3.51$ is estimated for the resonant-cavity SPP mode in the active medium of the cavity.
For ideal phase-matching, a mode-index value of $3.0$ is required, hence the phased-matching 
condition is not satisfied for this QCL. Consequently, the measured FWHM of the far-field radiation
pattern $\sim 12^\circ\times 16^\circ$ for this QCL is not as good as that for phased-matched third-order
DFB QCLs with reduced effective mode index~\cite{kao:dfbpm,amanti:dfb,amanti:dfbcw}. Elongated and multi-lobed beam pattern for third-order DFB when refractive index is close
to $3.6$ is observed in Ref.~\cite{amanti:dfb}. The peak output power for the  third-order DFB QCL without phase matching is $\sim 0.45~\rmmW$ at $78~\rmK$,
which is about three times less than that obtained for the antenna-feedback QCL with similar dimensions lasing at similar frequency.
The output power slope efficiencies at $78~\rmK$ are $4~$mW/A and $13~$mW/A respectively for the third-order DFB QCL and antenna-feedback QCL respectively.
The optical power was measured without cone collecting optics.
The relative differences in the measured optical power are in good agreement with the radiative losses estimated via 2D FEM simulations for similar
cavity dimensions as in Fig.~2. The maximum operating temperature of this QCL is almost similar to that of the antenna-feedback
QCL, which suggests that radiative loss is relatively a small fraction of the overall waveguide loss 
in both types of DFB QCL cavities. This also suggests that there is a scope for enhancing radiative
efficiency to increase optical power output without causing a significant degradation in temperature
performance of terahertz QCLs with antenna-feedback. Calculation of surface-outcoupling efficiency as in Ref.~\cite{boyle:high} is outside the scope of this work due to large uncertainty in the absolute loss contribution of metal layers, absorbing boundaries, as well as the active region itself. 
The optimized phase-matched third-order DFB ~\cite{amanti:dfbcw} with bound-to-continuum design shows lower current-density and higher efficiency without absorbing regions. Since the feedback of laser mode is achieved by SPP mode propagating on the surface for antenna-feedback, technique of covering waveguide sidewall with metal~\cite{kumar:surfemit} or lateral corrugated grating~\cite{amanti:dfbcw}, which does not absorb effective outcoupled laser radiation from antenna-feedback, can replace lateral absorber layer here and will further enhance the outcoupling efficiency. L-I characteristics at different heat-sink temperature of a representative terahertz QCL with \FPerot\ (FP) cavity is shown in Fig.~\ref{figS3}. Exposed highly-doped lossy contact layer serving as absorber introduces $\sim 15~\rmK$ temperature degradation for DFB cavity. Because sub-wavelength mode confinement in FP cavities results in highly divergent output beams of THz QCLs with parallel-plate metallic cavities, output power is measured with a Winston cone and the detector was placed adjacent to the cryostat window. Diameter of the circular opening of cone  is $\sim 1.9~\rmmm$. Frequency of gain medium covers $\sim 2.8~\thz$ to $\sim 3.2~\thz$ at $\sim 78~\rmK$. 

\section{Analysis of antenna-feedback scheme by 2D and 3D simulations}
\label{analysis}

\begin{figure}[htbp]
\centering
\includegraphics[width=3.2in]{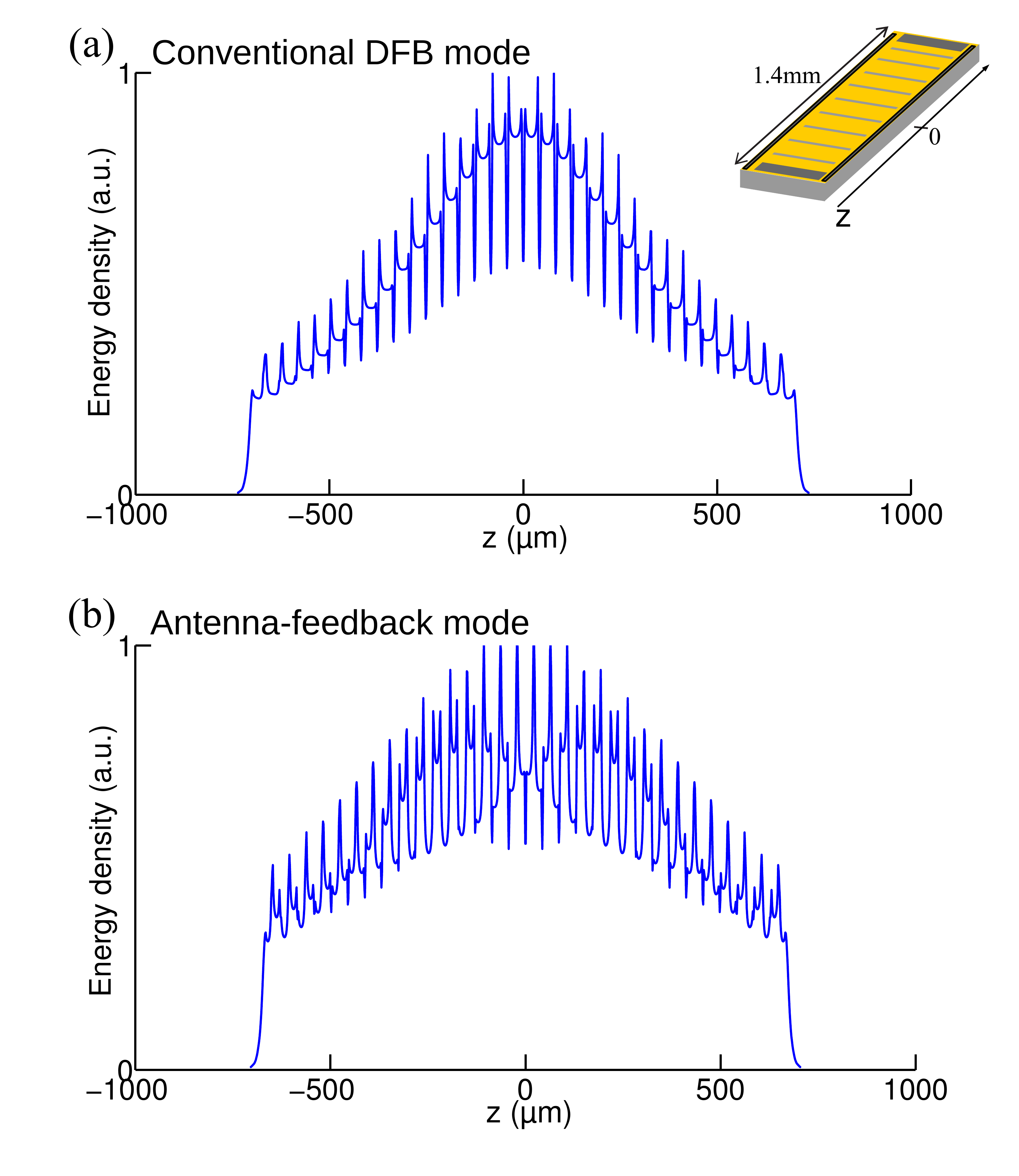}
\caption{
Energy-density profiles along the length of the metallic cavities (in $z$ direction) for the band-edge DFB modes for (a) third-order DFB cavity, and (b) cavity with antenna-feedback
respectively.
}
\label{figS4}
\end{figure}

Figure~\ref{figS4} shows energy-density profile for the resonant band-edge modes of the cavities with third-order DFB and antenna-feedback schemes respectively, along the entire length of the cavity ($\sim 1.4~$mm). Energy-density is calculated and summed up along the cavity's height ($10~\um$) from 2D simulations (that effectively model cavities of infinite width). The eigenmode spectra and electric-field distributions for the cavities are shown in Fig.~2. Both third-order DFB and antenna-feedback schemes show non-uniform envelope shapes, which gradually decrease from the center to the end-facets of cavity and provide indication of typical DFB action due to coupling of propagating waves along the length of cavity as per the corresponding schematics in Fig.~1.

\begin{figure}[htbp]
\centering
\includegraphics[width=3.5in]{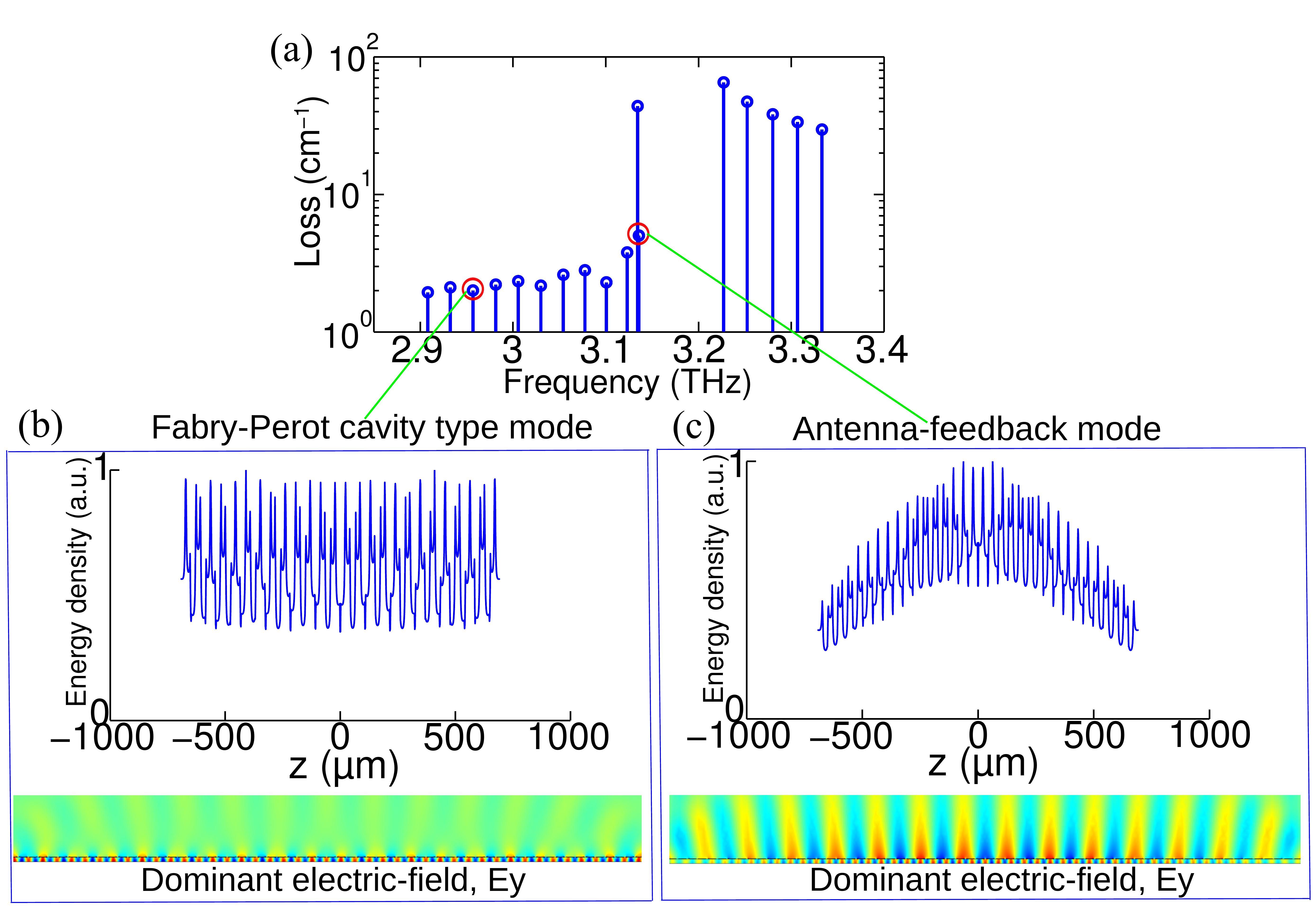}
\caption{
{\bf Role of longitudinal absorbing boundaries in resonant-cavities with antenna-feedback.}
(a) Eigenmode spectrum of QCL metallic cavity with antenna-feedback gratings in the top metallic layer, but without longitudinal absorbing boundaries at the two ends of the cavity. The dimensions of the cavity are otherwise similar to that for the result in Fig.~2(b). (b) Energy density profile along the length of the $\sim 1.4~\rmmm$ long cavity and dominant electric field $E_y$ of a \FPerot\ cavity type mode that exists sufficiently away from the photonic bandgap. (c) Energy density profile and electric-field profile along the length of the cavity for the ``desired'' antenna-feedback mode that exists at the edge of the photonic bandgap. The $z$ direction is along the length of laser ridge, as marked in Fig.~\ref{figS4}. 
}
\label{figS5}
\end{figure}

A thin highly-doped GaAs layer below the top-metal cladding was left unetched at the regions close to longitudinal facets, serving as the longitudinal absorbing boundary to ensure the excitation of the desired antenna-feedback mode as the lowest-loss lasing mode. Firstly, such absorbing boundaries  reduce the end-facet reflectivities that strengthens the distributed-feedback coupling for the antenna-feedback scheme. Secondly, the longitudinal absorber boundaries play an important role in providing the necessary mechanism for mode-discrimination, to selectively excite the desired band-edge antenna-feedback mode for lasing. Figure~\ref{figS5}(a) shows that eigenmode spectrum of antenna-feedback scheme without longitudinal absorbing boundaries. There exist the Fabry-Perot type modes with uniform energy density profile along the whole length of laser cavity and large electric field distribution close to end facets of the cavity, as shown in Fig.~\ref{figS5}(b). However, the loss of such modes is small due to highly reflective end-facets. The desired antenna-feedback mode, located at the lower bandedge, shows non-uniform energy-density profile along the length of cavities as an indicator of DFB action and strong coupling between plasmonic mode on top of metallic layer and guided mode in the active core, as seen from Fig.~\ref{figS5}(c). However, its radiative loss is higher due to the improved and large radiative field of antenna-feedback mode and hence, it cannot be excited for lasing in the absence of longitudinal absorbing boundaries. When longitudinal absorbing boundaries are introduced, the relative loss of the modes with large intensity closer to longitudinal ends of cavity, which are further away from the photonic bandgap, will be selectively and significantly increased. Therefore, the antenna-feedback mode is excited for lasing as the robust lowest-loss mode as seen from the eigenmode spectrum in Fig.~2(b).

\begin{figure}[htbp]
\centering
\includegraphics[width=3.3in]{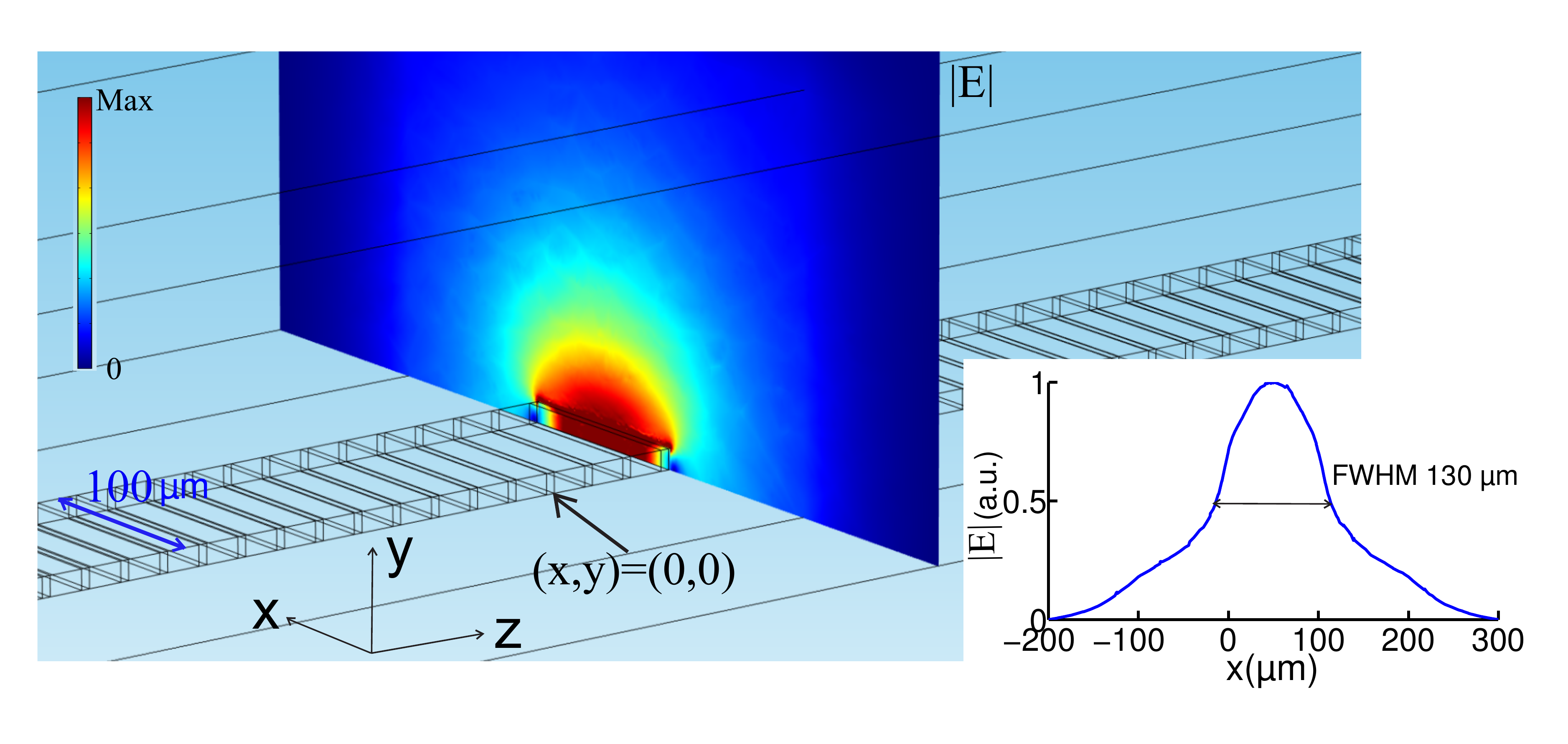}
\caption{
{\bf Electric-field profile of the SPP mode in surrounding medium of the plasmonic laser's cavity with antenna-feedback gratings from 3D simulation.}
A parallel-plate metal cavity with GaAs medium of $100~\um$ width, $10~\um$ thickness, and $1.4~\rmmm$ length is simulated as in Fig.~\ref{figS1}. The frequency of the band-edge resonant-cavity mode is $\sim 3.15~\thz$, which is the desired ``antenna-feedback'' mode. Inset: Mode shape along lateral $x$ dimension of cavity at a distance of $10~\um$ above the top metal cladding; the FWHM of the lateral electric-field profile is $\sim 130~\um$. 
}
\label{figS6}
\end{figure}

Figure~\ref{figS6} shows lateral mode shape of the antenna-feedback mode at approximately the longitudinal center of the cavity from 3D simulation. The SPP mode in surrounding medium extends in both lateral and vertical directions significantly, and the extent of the mode in lateral ($x$) dimension is greater than the width of the cavity itself. The FHWM of the mode shape along lateral ($x$) dimension of cavity at a distance of $\sim 10~\um$ above top gratings (picked arbitrarily to keep it slightly away from location where finite-element mesh in simulation changes abruptly) is $\sim 130~\um$. This large effective wavefront created on top of the waveguide leads to a narrow far-field, i.e. a narrow ``element-factor'' in terminology of the radiation from a phased-array antenna. While the narrowing of ``array-factor'' due to constructive interference of radiation from all apertures in end-fire direction is also at play; it is not the dominant effect to cause narrow beaming for terahertz QCLs in this work for short-length cavities ($<2~\rmmm$). In contrast, for third-order DFB QCLs with phase-matching, beam narrowing is primarily due to the narrow array-factor only, which is why very long cavities ($<5~\rmmm$) are needed to achieve beam divergence values of less than $\sim 10^\circ$~\cite{kao:dfbpm}.

\begin{figure}[htbp]
\centering
\includegraphics[width=3.3in]{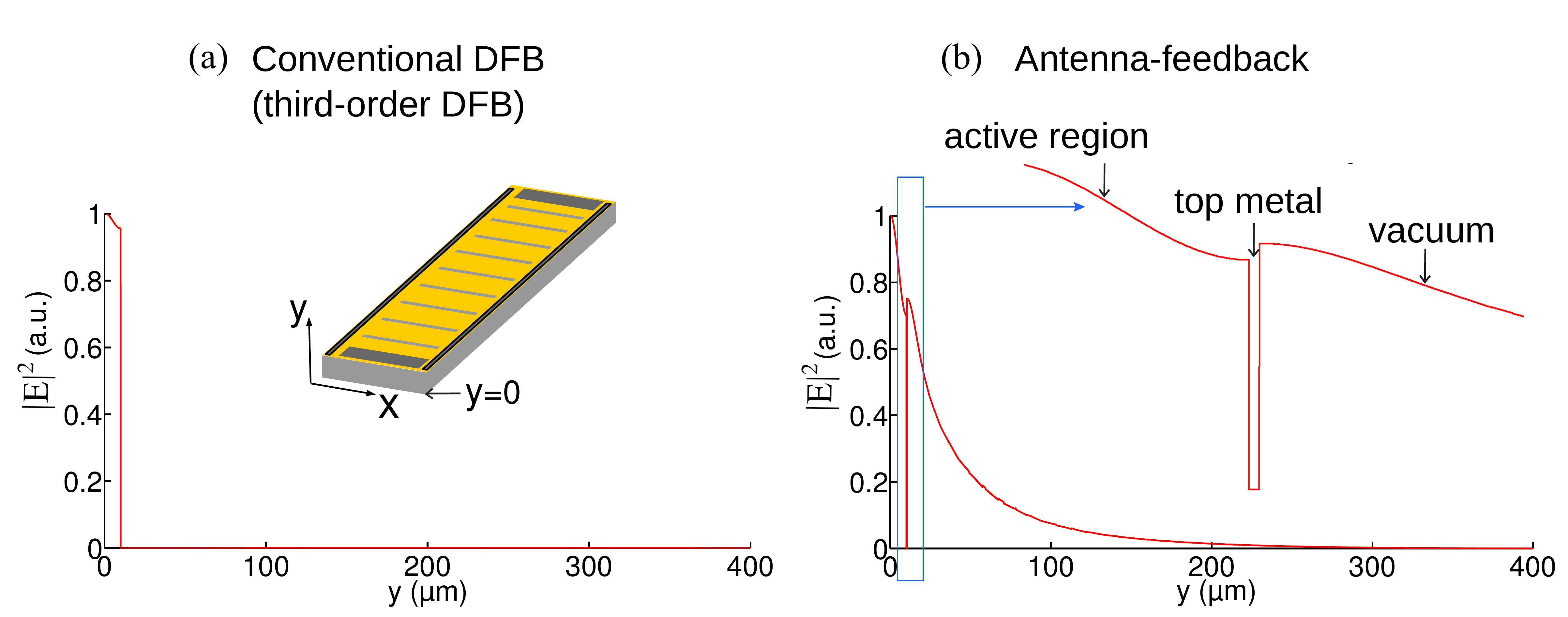}
\caption{
{\bf Comparison of electric-field squared modulus
$|E|^2$ profile in conventional DFB (third-order DFB as an example) and antenna-feedback scheme.}
(a) $|E|^2$ profile for third-order DFB cavity plotted along perpendicular $y$ direction (cavity's height direction). (b) $|E|^2$ profile for antenna-feedback scheme with an expanded view close to top metal cladding.
}
\label{figS7}
\end{figure}

Figure~\ref{figS7} shows the electric-field squared modulus $|E|^2$ profile inside the cavity and above the metal cladding for both conventional DFB and antenna-feedback scheme from 2D simulations. Layer sequence follows active region with $10~\um$ thickness, top metal layer with $400~$nm thickness (lossless metal) and air region. At the location above metal cladding, the mode-intensity in conventional DFB immediately decays to zero. In contrast, for antenna-feedback scheme, the sub-wavelength plasmonic resonant-optical mode inside the cavity is phased-locked with a SPP mode that travels on top of the metal cladding outside the cavity. The SPP wave is relatively tightly bound to the top metal cladding and decays in the length of the order of the wavelength in vacuum. The decay length is independent of the loss in metal or thickness in metal or various duty cycles of grating (which was verified via finite-element simulations). The establishment of a hybrid SPP mode on top of the cavity in vacuum is unique in antenna-feedback scheme. In a third-order DFB terahertz QCL, even with effective mode index getting close to $3.0$, the feedback mechanism remains same as that in conventional DFB lasers, i.e. distributed-feedback couples propagating modes inside the cavity without involving the propagating SPP waves in the surrounding medium. Correspondingly, no standing SPP wave (with large amplitude) is established in the surrounding medium in a third-order DFB QCL's cavity as it does in antenna-feedback scheme. This could be verified by simulating a terahertz QCL cavity structure with third-order DFB gratings, but with the refractive index of the active region arbitrarily set to a value of $\sim 3.0$ to satisfy the so-called ``phase-matching'' condition in simulation.

\end{document}